\documentclass[a4paper, 11pt]{article}
\pdfoutput=1
\usepackage{jheppub}
\makeatletter
\def\@fpheader{}
\makeatother

\usepackage{graphicx,wrapfig,float,slashed}
\usepackage{amsmath,amssymb,dsfont,epsfig,graphicx}
\usepackage{mathtools} % loads amsmath
\usepackage{multirow}
\usepackage{blkarray}
\usepackage{epstopdf}
\usepackage{ragged2e}
\usepackage[utf8]{inputenc} 
\usepackage{cancel}
\usepackage[usenames,dvipsnames,table]{xcolor}
\usepackage[toc]{appendix}
\usepackage[normalem]{ulem}
\usepackage{enumitem}
\usepackage[font=small,skip=1pt]{caption}
\allowdisplaybreaks
\usepackage{pifont}
\usepackage{subcaption}
\usepackage{footnote}

\newcommand{\nc}{\newcommand}
\nc{\non}{\nonumber}
\nc{\hc}{\hbox {h.c.}}
\nc{\noi}{\noindent}
\nc{\barx}{\bar{x}}
\nc{\pbarn}{\;\hbox {pb}}
\nc{\fbarn}{\;\hbox {fb}}

\nc{\hsp}{\hspace{0.5cm}}
\nc{\lsp}{\hspace{1cm}}
\nc{\Lsp}{\hspace{2cm}}
\nc{\LLsp}{\lsp\lsp}
\nc{\lra}{\longrightarrow}
\nc{\p}{\prime}
\nc{\sgn}{\text{sgn}}
\nc{\tr}{\text{Tr}}
\nc{\ph}{\varphi}
\nc{\op}{{\cal O}}
\nc{\cL}{{\cal L}}
\nc{\cU}{{\mathcal U}}
\nc{\cD}{{\mathcal D}}
\nc{\cQ}{{\mathcal Q}}
\nc{\cT}{{\mathcal T}}
\nc{\what}{\widehat}
\nc{\vmol}{v_{\text{M\o l}}}

\nc{\beq}{\begin{equation}}  \nc{\eeq}{\end{equation}}
\nc{\bea}{\begin{eqnarray}}  \nc{\eea}{\end{eqnarray}}
\nc{\baa}{\begin{array}}     \nc{\eaa}{\end{array}}
\nc{\bit}{\begin{itemize}}   \nc{\eit}{\end{itemize}}
\nc{\ben}{\begin{enumerate}} \nc{\een}{\end{enumerate}}
\nc{\bce}{\begin{center}}    \nc{\ece}{\end{center}}
\nc{\bpm}{\begin{pmatrix}}   \nc{\epm}{\end{pmatrix}}
\nc{\bvt}{\begin{verbatim}}  \nc{\evt}{\end{verbatim}}

\def\lsim{\mathrel{\raise.3ex\hbox{$<$\kern-.75em\lower1ex\hbox{$\sim$}}}}
\def\gsim{\mathrel{\raise.3ex\hbox{$>$\kern-.75em\lower1ex\hbox{$\sim$}}}}

\def\udots{\mathinner{\mkern1mu\raise1pt\vbox{\kern7pt\hbox{.}}\mkern2mu\raise4pt\hbox{.}\mkern2mu\raise7pt\hbox{.}\mkern1mu}}

\def\mev{\;\hbox{MeV}}
\def\gev{\;\hbox{GeV}}
\def\tev{\;\hbox{TeV}}

\definecolor{agray}{rgb}{0.95, 0.95, 0.99}

\def\app#1{Appendix~\ref{#1}}
\def\eq#1{Eq.~(\ref{#1})}
\def\fig#1{Fig.~\ref{#1}}
\def\sec#1{Sec.~\ref{#1}}

\begin{document}
%%%%%%%%%%%%%%%%%%%%%%%%%

%%%%%%%%%%%%%%%%%%%%%%%%%%%%%%%%%%%%%%%%%%%%%%%%%%%%%%%%%%%%%%%%%%%%%%%%%%%%%%%%%%%%%%%%%%%%%%%%%%%%%%%%%%%%%%%%%%%
\title{A Minimal Model for Neutral Naturalness and pseudo-Nambu-Goldstone Dark Matter}
\author[1]{Aqeel Ahmed,}
\emailAdd{aqeel.ahmed@vub.be}
\author[1]{Saereh Najjari,}
\emailAdd{saereh.najjari@vub.be}
\author[2]{and Christopher B. Verhaaren}
\emailAdd{cverhaar@uci.edu}
\affiliation[1]{Theoretische Natuurkunde \& IIHE/ELEM, \\ Vrije Universiteit Brussel, Pleinlaan 2, 1050 Brussels, Belgium}
\affiliation[2]{Department of Physics and Astronomy,\\ University of California, Irvine, USA}

%%%%%%%%%%%%
\abstract{
We outline a scenario where both the Higgs and a complex scalar dark matter candidate arise as the pseudo-Nambu-Goldstone bosons of breaking a global $SO(7)$ symmetry to $SO(6)$. 
The novelty of our construction is that the symmetry partners of the Standard Model top-quark are charged under a hidden color group and not the usual $SU(3)_c$. 
Consequently, the scale of spontaneous symmetry breaking and the masses of the top partners can be significantly lower than those with colored top partners. 
Taking these scales to be lower at once makes the model more natural and also reduces the induced non-derivative coupling between the Higgs and the dark matter. 
Indeed, natural realizations of this construction describe simple thermal WIMP dark matter which is stable under a global $U(1)_D$ symmetry. 
We show how the Large Hadron Collider along with current and next generation dark matter experiments will explore the most natural manifestations of this framework.
}
%%%%%%%%%%%%

\preprint{UCI-TR-2020-04}
\arxivnumber{2003.08947}
%%%%%%%%%%%%%
%\toccontinuoustrue 

\flushbottom
\maketitle
\flushbottom
%%%%%%%%%%%%

%%%%%%%%%%%%%%%%%%%%%%%%%%%%%%%%
\section{Introduction} 
\label{introduction}
%%%%%%%%%%%%%%%%%%%%%%%%%%%%%%%%%%%

The Standard Model (SM) of particle physics has great agreement with experiment, however it cannot be the complete theory of nature. One of the most pressing theoretical problems within the SM is the hierarchy between the weak and Planck scales. Both composite Higgs models and constructions which protect the Higgs mass through a new symmetry predict new particles or states with masses at or below the TeV scale.

Beyond this theoretical puzzle, there is overwhelming experimental evidence for dark matter (DM) which also points to new particles and interactions beyond the SM. While there is a vast and varied spectrum of possible DM candidates, weakly interacting massive particles (WIMPs) are perhaps the most theoretically compelling. This is especially the case when viewed through the lens of the hierarchy problem. Then the DM can naturally obtain a weak scale mass and couplings, providing the observed DM density through thermal freeze-out. 

However, both symmetry based explanations of Higgs naturalness and thermal WIMPs have become increasingly constrained by experiment. Searches at the Large Hadron Collider (LHC) have pushed the limits on the colored symmetry partners of SM quarks to the TeV scale. At the same time a host of direct detection experiments are driving the limits on WIMP DM cross sections toward the so-called neutrino floor. With the severity of these constraints many new and interesting ideas for both Higgs naturalness and DM have been explored.

Years before the Higgs was discovered~\cite{Aad:2012tfa,Chatrchyan:2012xdj}, it was pointed out that if the Higgs mass parameter is insensitive to high scales because of a new symmetry, the symmetry partners of the SM quarks do not need to carry SM color~\cite{Chacko:2005pe,Barbieri:2005ri,Burdman:2006tz,Poland:2008ev,Cai:2008au}. Since the discovery of colored symmetry partners to SM quarks has not followed the discovery of the Higgs, more realizations of color neutral naturalness have been explored~\cite{Craig:2014aea,Craig:2014roa,Batell:2015aha,Serra:2017poj,Csaki:2017jby,Cohen:2018mgv,Cheng:2018gvu,Dillon:2018wye,Xu:2018ofw,Serra:2019omd}. Connecting DM to neutral naturalness began with the Dark Top~\cite{Poland:2008ev}, and has flourished in the context of twin Higgs models~\cite{Garcia:2015loa,Craig:2015xla,Garcia:2015toa,Farina:2015uea,Freytsis:2016dgf,Farina:2016ndq,Barbieri:2016zxn,Barbieri:2017opf,Hochberg:2018vdo,Cheng:2018vaj,Terning:2019hgj,Koren:2019iuv,Badziak:2019zys}.

In twin Higgs models, the Higgs is a pseudo-Nambu-Goldstone boson (pNGB) of a global $SO(8)$ symmetry breaking to $SO(7)$. The variety in DM candidates typically comes not from the symmetry breaking structure, but by making particular assumptions about the particle content in the twin sector. Other neutral naturalness pNGB constructions~\cite{Cai:2008au,Serra:2017poj,Csaki:2017jby,Xu:2018ofw,Serra:2019omd} employ smaller symmetry groups, but this move toward minimality makes it more difficult to accommodate simple DM candidates. 

However, as demonstrated in~\cite{Balkin:2017aep,Balkin:2018tma}, the six pNGBs that spring from $SO(7)/SO(6)$ can be associated with the Higgs doublet (respecting the custodial symmetry) along with a complex scalar DM stabilized by a global $U(1)_D$~\footnote{A coset like $SO(6)/SO(5)$ leads to five pNGBs which comprise the Higgs doublet and a real scalar field which can be a dark matter candidate~\cite{Frigerio:2012uc}, however the stability of DM requires an additional dark pairity.}. The mass of the DM and its couplings to the Higgs are determined by the symmetry breaking structure and the low energy fields that transform under the symmetry. This necessarily includes the top quark for the model to address the hierarchy problem. As a consequence, the collider bounds on colored top partners lead to couplings between the Higgs and the DM that are near or beyond the experimental limits~\cite{Balkin:2017aep,DaRold:2019ccj}.

In the following section we construct a neutral natural version of the $SO(7)/SO(6)$ symmetry breaking pattern. As in other neutral natural models, the quark symmetry partners are charged under a hidden color group $SU(3)_{\what{c}}$ which is related to SM color by a discrete $\mathbb{Z}_2$ symmetry in the UV. This means they can be much lighter, allowing for additional freedom in the Higgs non-derivative coupling to the DM. These SM color neutral top partners are electroweak charged and break the DM shift symmetry, generating the DM potential. Thus, more natural top partner masses can lead to Higgs portal direct detection signals that may not be fully explored until the next generation of dark matter experiments. However, we do find that nearly all natural parameter choices lie above the neutrino floor. 

In addition, the new fields related to the top quark exhibit quirky~\cite{Kang:2008ea} dynamics. These less studied particles can be discovered at the LHC, providing a complementary probe of the model. In Sec.~\ref{sec:pheno} we outline the most promising collider searches, including both prompt and displaced signals. We find that the LHC already bounds the quirks with $U(1)_D$ charges. Because these particles determine the coupling between the Higgs and the dark matter, these collider bounds immediately inform the sensitivity of dark matter experiments to the pNGB WIMPs. We also calculate the corrections to the electroweak precision tests (EWPT) due to the presence of the new electroweak charged states.

In Sec.~\ref{sec:DMpheno} we discuss the DM phenomenology, showing which parameter values lead to the correct thermal relic density and elucidate how direct and indirect searches probe the model. We find collider searches and direct detection experiments provide complementary probes, both delving into the natural parameter space along different directions in parameter space. While current limits allow versions of this framework with $\sim 10$\% tuning, next generation searches should be able to discover the quirks or DM, often in multiple channels, down to $\sim1$\% tuning. Following our conclusions, in Sec.~\ref{s.conclusion}, we include two appendices to provide details relating to the work. 

%%%%%%%%%%%%%%%%%%%%%%%%%%%%%%%%%%%
\section{Neutral Naturalness from $SO(7)/SO(6)$\label{s.model}}
%%%%%%%%%%%%%%%%%%%%%%%%%%%%%%%%%%%
In this section, we describe a neutral naturalness model which includes the Higgs doublet and a complex scalar DM candidate as pNGBs. This model is related to that of the Refs.~\cite{Balkin:2017aep,Balkin:2018tma}, but crucially has color neutral top partners. The global symmetry structure is $SO(7)\times U(6)$, where $U(6)\simeq SU(6)\times U(1)_X\supset SU(3)_c\times SU(3)_{\what{c}}\times U(1)_X$ includes the SM color group as well as a hidden sector color denoted $SU(3)_{\what{c}}$. This hidden color symmetry is assumed to be related to the SM color group by a discrete $\mathbb{Z}_2$ symmetry in the UV. While this discrete symmetry is broken at lower energies, this symmetry ensures that the Yukawa coupling between the Higgs and top sector runs similarly in both theories. Though this is a two-loop effect, it has been shown that, because the strong coupling is somewhat large, it can significantly affect the fine-tuning of neutral natural models~\cite{Craig:2015pha}. 

The additional $U(1)_X$ ensures SM fields have their measured hypercharges. At some scale~$f$ the global $SO(7)$ symmetry is broken to $SO(6)\supset SO(4)_{C}\times SO(2)_D\cong SU(2)_L\times SU(2)_R\times U(1)_D$. Here the $SO(4)_C\simeq SU(2)_L\times SU(2)_R$ is the familiar custodial symmetry with $SU(2)_L$ being the usual SM weak gauge group and $SO(2)_D\!=\!U(1)_D$ is the global symmetry that stabilizes the DM. This construction also breaks the DM's shift symmetry in a new way. In particular, through color neutral vector-like quarks in addition to the color neutral top partners. As we see below, the DM mass and its non-derivative couplings are proportional to the masses of these color neutral vector-like quarks. 

%%%%%%%%%%%%%%%%
\subsection{The Gauge Sector}
We begin with the interactions amongst the NGBs and the gauge fields. The NGB fields can be parameterized nonlinearly as
\beq
\Sigma= e^{-i\Pi/f}\Sigma_0, \lsp {\rm with}\lsp \Pi=\sqrt2 \pi_{\hat a} T^{\hat a},
\eeq
where $\Sigma_0=(0,0,0,0,0,0,f)^T$ and $T^{\hat{a}}$ are the broken generators of $SO(7)$ with $\hat{a}=1,\cdots,6$, see Appendix~\ref{app:gens} for details. We immediately find
\begin{equation}
\Sigma=\frac{f}{|\pi|}\sin\Big(\frac{|\pi|}{f}\Big)\left(\pi_1,\pi_2,\pi_3,\pi_4,\pi_5,\pi_6,|\pi|\cot\Big(\frac{|\pi|}{f}\Big)\right)^T,
\end{equation}
where $|\pi|\equiv \sqrt{(\pi^{\hat a})^2} $. We can then write the leading order NGB Lagrangian as
\beq
\mathcal{L}_\text{NGB}=\frac{1}{2}\left(D_\mu\Sigma \right)^TD^\mu\Sigma,
\eeq
 where the covariant derivative is given by
 \beq
 D_\mu=\partial_\mu-igW^a_\mu T^a_L-ig'B_\mu\left(T^3_R+X \right).
 \eeq
 Note that the electric charge of fields is defined by $Q=T^3_L+T^3_R+X$, or the hypercharge is defined as $Y=T^3_R+X$.
 
 The first four NGBs are related to the usual Higgs doublet $H=(h^+, h^0)^T$ by
\beq
\left( \pi_1,\pi_2,\pi_3,\pi_4\right)=\left(-i\frac{h^+-h^{+\ast}}{\sqrt{2}},\frac{h^++h^{+\ast}}{\sqrt{2}},i\frac{h^0-h^{0\ast}}{\sqrt{2}},\frac{h^0+h^{0\ast}}{\sqrt{2}} \right).
\eeq
In unitary gauge when $h^+=0$ and $h^0=\overline{h}/\sqrt{2}$ we have
\beq
\left( \pi_1,\pi_2,\pi_3,\pi_4,\pi_5,\pi_6\right)=\left(0,0,0,\overline{h},\sqrt{2}\,\text{Im}\chi,\sqrt{2}\,\text{Re}\chi\right),
\eeq
where we have defined $\chi=(\pi_6+i\pi_5)/\sqrt{2}$ as a complex scalar which is our DM candidate. It is convenient to make the field redefinition~\cite{Gripaios:2009pe},
\beq
\sin\left(\frac{|\pi|}{f}\right)\frac{\pi^a}{|\pi|}\to\frac{\pi^a}{f}.
\eeq
We can then write NGB field as
\beq
\Sigma=\left(0,0,0,\overline{h},\sqrt{2}\,\text{Im}\chi,\sqrt{2}\,\text{Re}\chi,\sqrt{f^2-\overline{h}^2-2|\chi^2|}\, \right),
\eeq
in unitary gauge. The NGB Lagrangian has the leading order terms
\begin{align}
\mathcal{L}_\text{NGB}=&\frac12\left(\partial_\mu\overline{h}\right)^2+\left|\partial_\mu\chi \right|^2+\frac12\frac{\left(\overline{h}\partial_\mu\overline{h}+\chi^\ast\partial_\mu\chi+\chi\partial_\mu\chi^\ast \right)^2}{f^2-\overline{h}^2-2|\chi^2|}\nonumber\\
&+\frac{\overline{h}^2}{4}\left[g^2W^{+\mu}W^-_\mu+\frac12\left(gW^3_\mu-g'B_\mu \right)^2 \right].
\end{align}
When $\overline{h}$ gets a vacuum expectation value (VEV) of $v\approx246$ GeV we write 
\beq
\overline{h}=v+c_vh, \ \ \ \ \text{with} \ \ \ \ c_v\equiv\sqrt{1-\tfrac{v^2}{f^2}},\label{e.csvoverf}
\eeq
to ensure that $h$ is canonically normalized. We also define $s_v\equiv v/f$.

%%%%%%%%%%%%%%%
\subsection{The Quark Sector}
The quark fields include particles charged under both SM color and the hidden color group. In terms of $SO(7)$ and $SU(6)$ representations we have the left- and right-handed quarks as ${\cal Q}_L=({\bf  7 , 6})$ and ${\cal T}_R=({\bf 1, 6})$. These can be split up schematically in terms of fields in the ${\bf 3}$ of their respective color groups
\beq
\mathcal{Q}_L=(Q_L,\widehat{Q}_L), \ \ \ \ {\cal T}_R=(t_R,\widehat{T}_R),
\eeq
where we have put a hat on fields charged under the hidden color group. More explicitly we write out the low lying left-handed fields as
\beq
\sqrt{2}{\cal Q}_L=\begin{blockarray}{lcc}
& SU(3)_c & SU(3)_{\what{c}} \\
\begin{block}{l(c|c)}
\multirow{4}{*}{$SO(4)_C$} & ib_L& i\widehat{b}_L \\ 
  & b_L & \widehat{b}_L \\ 
 & it_L & i\widehat{t}_L \\ 
  & -t_L & -\widehat{t}_L \\  \BAhhline{&--}
 \multirow{2}{*}{$U(1)_D$} & 0 &\displaystyle i\widehat{X}_L-i\widehat{Y}_L \\
 & 0 & \widehat{X}_L+\widehat{Y}_L  \\ \BAhhline{&--}
 & 0 &\displaystyle\sqrt{2} \widehat{T}_L \\
\end{block}
\end{blockarray}\;\; ,
\eeq
where $q_L=(t_L, b_L)^T$ is the usual $SU(2)_L$ quark doublet. This is similar in spirit to Refs.~\cite{Balkin:2017aep,Serra:2017poj,Csaki:2017jby} which use incomplete quark multiplets. One can imagine the other fields lifted out of the low energy spectrum by vector-like masses, or as in extra dimensional models~\cite{Burdman:2006tz,Cai:2008au} that the boundary conditions of the bulk fields are such that the zero modes vanish. In order to obtain the correct hypercharge for $t_L$, $t_R$, and $b_L$, both $\mathcal{Q}_L$ and $\mathcal{T}_R$ have a $U(1)_X$ charge of 2/3. The Yukawa coupling term $\overline{\mathcal{Q}}_L\Sigma\mathcal{T}_R$ then implies that the NGBs have zero $X$ charge, which in particular implies that $\chi$ has no SM gauge charges. 

\begin{figure} [t!]
\centering
\includegraphics[width=0.8\textwidth]{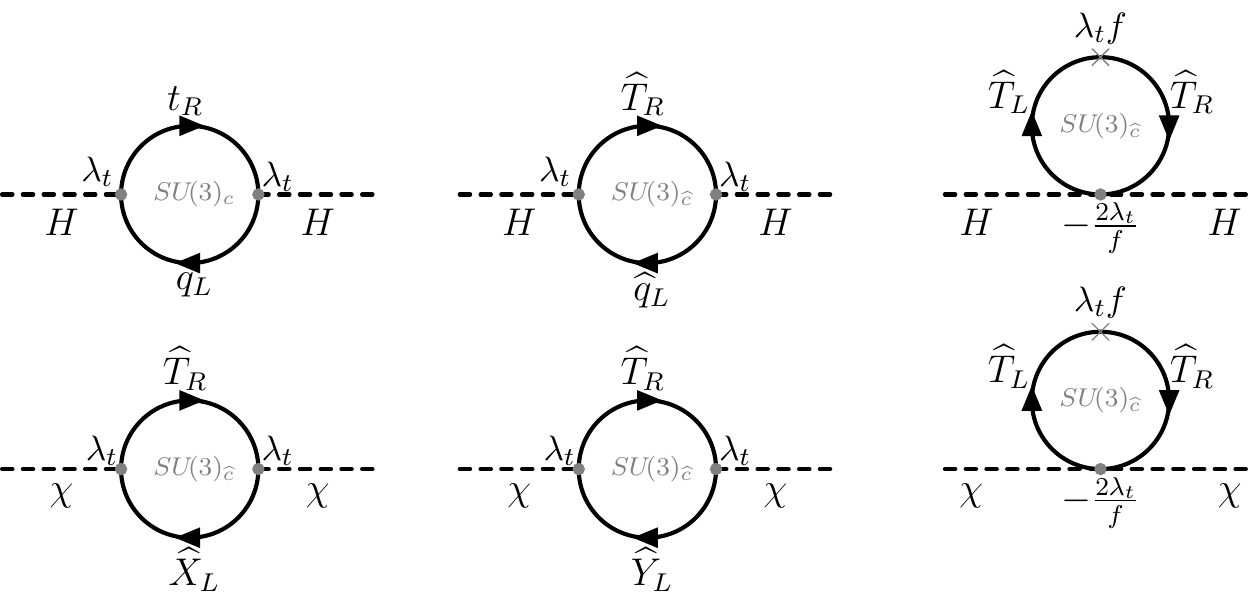} 
\caption{These Feynman diagrams show cancelation of quadratic divergences due to $SU(3)_c$ top quark and $SU(3)_{\what c}$ top quark for the SM-like Higgs $H$ (upper row) and the complex scalar $\chi$ (lower row).}
\label{fig:feyn_cancel}
\end{figure}

The top sector couplings follow from
\begin{align}
\mathcal{L}&\supset\lambda_t\overline{\mathcal{Q}}_L\Sigma\mathcal{T}_R+\text{h.c.}\label{e.TopSector}\\
&=\lambda_t\!\left[\!-\!\left(\overline{q}_Lt_R+\overline{\widehat{q}}_L\widehat{T}_R \right)\widetilde{H}\!+\!\chi^\ast\overline{\widehat{X}}_L\widehat{T}_R \!+\!\chi\overline{\widehat{Y}}_L\widehat{T}_R \!+\!f\Big(1-\frac{|H|^2+|\chi|^2}{f^2}+\ldots \Big)\overline{\widehat{T}}_L\widehat{T}_R\right]\!+\!\text{h.c.},\notag
\end{align}
where $q_L\!=\!(t_L, b_L)^T$ and $\what q_L\!=\!(\what t_L, \what b_L)^T$ are $SU(2)_L$ doublets and we have restored the eaten NGBs for the moment into the Higgs doublet $H$, and defined $\widetilde{H}=i\sigma^2H^\ast$. From these interactions we obtain the one-loop diagrams in Fig.~\ref{fig:feyn_cancel} relevant to the mass corrections for $H$ and $\chi$. The leading contributions from the top quark are doubled by $\widehat{q}_L$ interaction, but this combination is exactly cancelled by $\widehat{T}_L$. Like in~\cite{Xu:2018ofw}, the contributions from fields carrying SM color and those carrying hidden color are not equal. Note that the DM shift symmetry is broken by the SM color neutral top partner $\what T$ and $U(1)_D$ charged fermions $\what X_L,\what Y_L$.

The hidden color fields can be lifted through vector-like mass terms with new heavy states. We can write down the mass terms
\beq
\mathcal{L}_\text{vec mass}=-m_Q\overline{\widehat{q}}_L\widehat{q}_R-m_{X}\overline{\widehat{X}}_L\widehat{X}_R -m_{Y}\overline{\widehat{Y}}_L\widehat{Y}_R +\text{h.c.}~,
\eeq
where $\widehat{q}_R^{}=(\widehat{t}_R,\widehat{b}_R)^T$ is an $SU(2)_L$ doublet. We take these to be free parameters as we calculate the scalar potential.

%%%%%%%%%%%%%%%%
\subsection{The Scalar Potential \label{sec:ScalarPotential}}
%%%%%%%%%%%%%%%%
We are interested in the obtaining the potentials for both the Higgs and the DM. This is obtained from the Coleman-Weinberg (CW) potential~\cite{Coleman:1973jx}
\beq
V_\text{CW}=-\frac{N_c}{8\pi^2}\Lambda_\text{UV}^2\text{Tr}\mathcal{M}^2-\frac{N_c}{16\pi^2}\text{Tr}\left[\mathcal{M}^4\left(\ln\frac{\mathcal{M}^2}{\Lambda_\text{UV}^2}-\frac12 \right) \right],\label{e.CWpot}
\eeq
where $\mathcal{M}^2$ is the Dirac fermion mass squared matrices, with masses as functions of $\overline{h}$ and $\chi$. We note first that there is no quadratic sensitivity to to the cut off because $\text{Tr}\,\mathcal{M}^2$ is independent of the scalar fields. However, we do find logarithmic sensitivity because 
\beq
\text{Tr}\,\mathcal{M}^4=\frac{\lambda_t^4}{2}\overline{h}^4+\overline{h}^2\lambda_t^2\left(m_Q^2-\lambda_t^2 f^2 \right)+2\lambda_t^2|\chi|^2\left(m_X^2+m_{Y}^2 \right),
\eeq
where we have dropped field independent terms. 

Any remaining terms in the scalar potential, such as quartic mixing of $\overline{h}$ and $\chi$ or a $|\chi|^4$ term, are independent of $\Lambda_\text{UV}$ and so are robustly determined by the low energy physics. Clearly, in order for electroweak symmetry to break we need the Higgs mass parameter to be negative, so we require $f\lambda_t>m_Q$. From Eq.~\eqref{e.csvoverf} we see that Higgs couplings to SM fields will be reduced by $c_v$. As in other pNGB Higgs construction, this implies that $f$ exceeds $v$ by a factor of a few. As in~\cite{Xu:2018ofw} we find there must be a cancellation between independent terms ($m_Q$ and $\lambda_tf$) to obtain the correct Higgs mass. This motivates defining 
\beq
\lambda_t^2 f^2\equiv m_Q^2(1+\delta_m).
\eeq

For simplicity, in this work we take the vector-like masses of the DM sector to be equal
\beq
m_X=m_Y\equiv m_V.
\eeq
This mass scale is related to $m_Q$ by the ratio $r_Q=m^2_V/m^2_Q$. In this limit we find one of the dark fermion mass eigenstates is exactly $m_V$, while the others are determined by a cubic equation. We then find the scalar potential, which has the general form of
\beq
V(h,\chi)=\frac12\mu_h^2\overline{h}^2+\frac{\lambda_h}{4}\overline{h}^4+\mu^2_\chi|\chi|^2+\lambda_\chi|\chi|^4+\lambda_{h\chi}\overline{h}^2|\chi|^2~. \label{eq:eff_pot}
\eeq
The potential parameters are calculated from the CW potential in Eq.~\eqref{e.CWpot}. We find
\begin{align}
\mu_h^2&=\frac{3\lambda_t^2}{8\pi^2}\bigg[m_Q^2\ln\frac{\Lambda_\text{UV}^2}{m_Q^2}-\lambda_t^2 f^2\ln\frac{\Lambda_\text{UV}^2}{\lambda_t^2 f^2} +\frac{\lambda_t^2 f^2\,m_Q^2}{m_Q^2-\lambda_t^2 f^2}\ln\frac{\lambda_t^2 f^2}{m_Q^2}\bigg],\\
\lambda_h&=\frac{3\lambda_t^4}{16\pi^2}\bigg[\ln\frac{\Lambda_\text{UV}^4}{\lambda_t^2 f^2\frac{\lambda_t^2}{2}\overline{h}^2}-\frac12\frac{\big(3m_Q^2-\lambda_t^2 f^2 \big)^2}{\big(m_Q^2-\lambda_t^2 f^2\big)^2}-m_Q^4\frac{3m_Q^2-\lambda_t^2 f^2}{\big(m_Q^2-\lambda_t^2 f^2\big)^3}\ln\frac{\lambda_t^2 f^2}{m_Q^2} \bigg],\\
\mu_\chi^2&=\frac{3\lambda_t^2m_V^2}{4\pi^2}\bigg[\ln\frac{\Lambda_\text{UV}^2}{\lambda_t^2 f^2}+\frac{m_V^2}{m_V^2-\lambda_t^2f^2}\ln\frac{\lambda_t^2f^2}{m_V^2} \bigg],\\
\lambda_\chi&=-\frac{3\lambda_t^4m_V^4}{4\pi^2\big( m_V^2-\lambda_t^2f^2\big)^2}\bigg[ 2+\frac{m_V^2+\lambda_t^2f^2}{m_V^2-\lambda_t^2f^2}\ln\frac{\lambda_t^2f^2}{m_V^2}\bigg],\\
\lambda_{h\chi}&=-\frac{3\lambda_t^4m_V^2}{8\pi^2}\bigg[\frac{2m_Q^2-\lambda_t^2 f^2}{(m_Q^2-\lambda_t^2 f^2 )( m_V^2-\lambda_t^2 f^2)}+ \frac{m_V^2\big(2m_Q^2-m_V^2 \big)}{\big(m_V^2-\lambda_t^2f^2\big)^2(m_Q^2-m_V^2 )}\ln\frac{\lambda_t^2f^2}{m_V^2}\nonumber\\
&\hspace{2cm}- \frac{m_Q^4}{\big(m_Q^2-\lambda_t^2f^2\big)^2(m_Q^2-m_V^2 )}\ln\frac{\lambda_t^2f^2}{m_Q^2}\bigg].
\end{align}

We need the dark $U(1)_D$ to remain unbroken so that $\chi$ is stable. This means we are interested in vacua with $\langle\chi\rangle=0$ and $\langle \overline h\rangle=v$. With $\mu_h^2<0$ and $\mu_\chi^2>0$, this is the deepest vacuum as long as $\lambda_h\lambda_\chi<\lambda_{h\chi}^2$. However, when $\lambda_h\lambda_\chi>\lambda_{h\chi}^2$ the vacuum with $\langle \chi\rangle\neq0$ becomes a saddle point rather than a minimum, so the deepest stable vacuum still has $\langle\chi\rangle=0$.
In this case we find 
\beq
\mu_h^2=-\lambda_hv^2, \lsp m_h^2=2c_v^2\lambda_hv^2, \lsp  m_\chi^2=\mu_\chi^2+\lambda_{h\chi}v^2.
\eeq
Since we know $v\simeq 246$ GeV and the Higgs mass $m_h\simeq125$ GeV, therefore $\lambda_h\simeq0.13$ and $\mu_h\simeq 89$ GeV. The constraints on Higgs couplings (see Sec.~\ref{sec:pheno}) imply that $f\gtrsim 3v$, which means $\delta_m\ll 1$. It then makes sense to expand the potential terms to leading order in $\delta_m$. We find
\begin{align}
\mu_h^2&=-\frac{3\lambda_t^2}{8\pi^2}m_Q^2+\mathcal{O}(\delta_m), \lsp 
\lambda_h=\frac{3\lambda_t^4}{16\pi^2}\bigg[\frac23+\ln\frac{\Lambda_\text{UV}^4}{m_Q^2\frac{\lambda_t^2}{2}\overline{h}^2}+\mathcal{O}(\delta_m)\bigg],\\
\mu_\chi^2&=\frac{3\lambda_t^2m_V^2}{4\pi^2}\bigg[\ln\frac{\Lambda_\text{UV}^2}{m_Q^2}+\frac{r_Q}{1-r_Q}\ln r_Q+\mathcal{O}(\delta_m)\bigg],\label{e.chiMass}\\
\lambda_\chi&=\frac{3\lambda_t^4r_Q^2}{4\pi^2\left(1-r_Q\right)^2}\bigg[\frac{1+r_Q}{1-r_Q}\ln\frac{1}{r_Q}-2 +\mathcal{O}(\delta_m)\bigg],\\
\lambda_{h\chi}&=\frac{3\lambda_t^4r_Q}{8\pi^2(1-r_Q)^2}\bigg[\frac{3-r_Q}{2}-\frac{r_Q(2-r_Q)}{1-r_Q}\ln\frac{1}{r_Q} +\mathcal{O}(\delta_m)\bigg].		\label{eq:lamhchi}
\end{align}
Here we have taken $\ln\frac{\Lambda_\text{UV}^2}{m_Q^2}$ to be order one, as expected for a cutoff of a few TeV. 

The Higgs potential has logarithmic dependence on $\Lambda_\text{UV}$. This is similar to both the Twin Higgs~\cite{Contino:2017moj} and $SO(6)/SO(5)$ constructions~\cite{Serra:2017poj} where sizable UV contributions lead to the correct Higgs mass. In the limit of small $\delta_m$ and taking $\lambda_t^{\overline{\text{MS}}}=0.936$ we find $\mu^2_h\approx -(146)^2$ $\text{GeV}^2$ and $\lambda_h\approx0.13$ for $m_Q=800$ GeV and $\Lambda_\text{UV}=3$ TeV. These are similar to the SM listed above,  so we expect that a suitable UV completion, perhaps composite or holographic, can easily accommodate the measured Higgs mass.

\begin{figure} [t]
\centering
\includegraphics[width=0.8\textwidth]{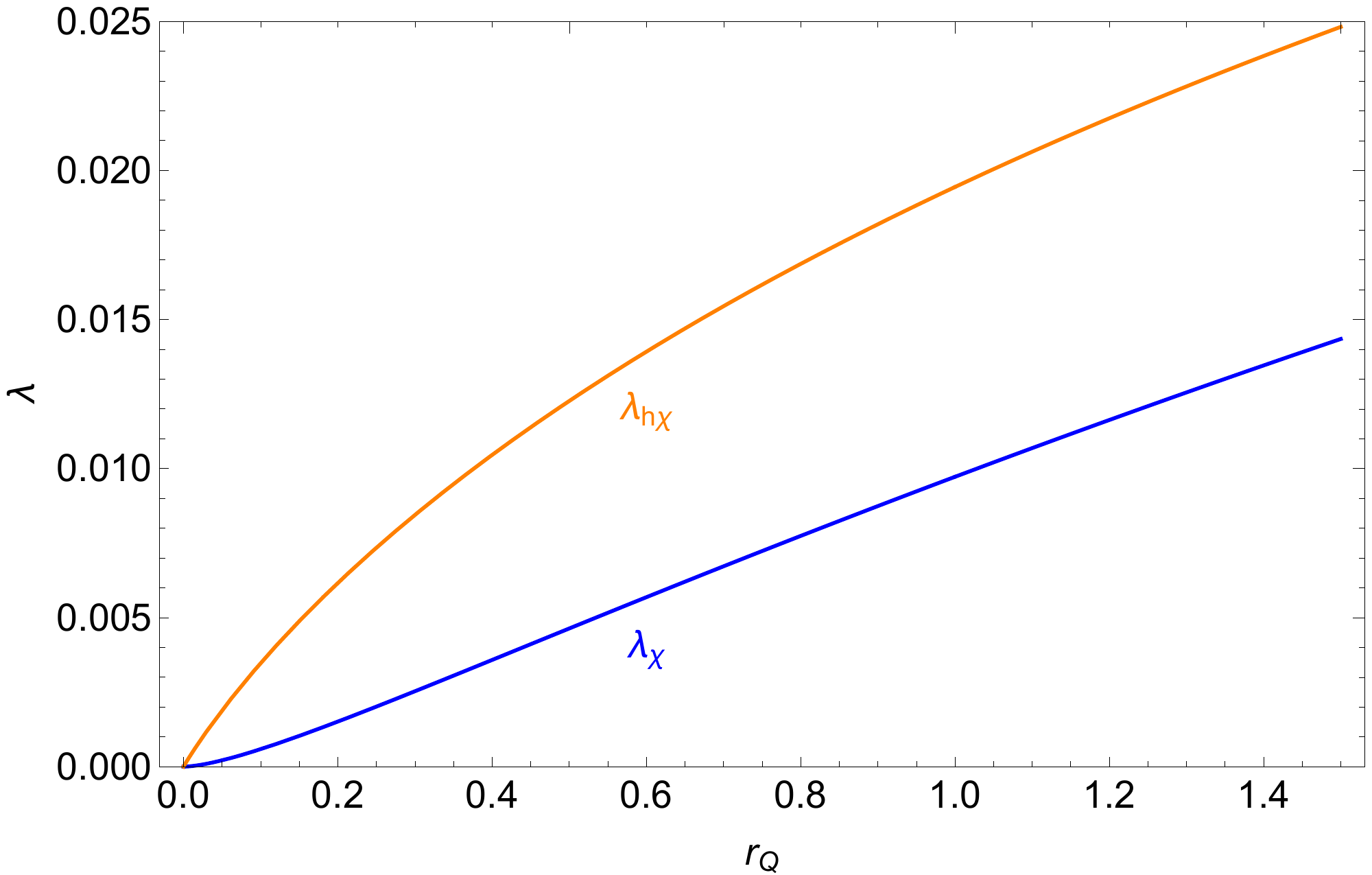} 
\caption{The quartic couplings involving the DM field $\chi$ as a function of $r_Q=m_V^2/m_Q^2$. We have neglected all terms of order $\delta_m$.}
\label{fig:quartics}
\end{figure}

At the same time the quartic couplings that involve $\chi$ are determined completely by the low-energy theory. Thus, we can make robust predictions about the DM without knowledge of the UV completion. In Fig.~\ref{fig:quartics} we see that these quartics are order $10^{-2}$ over a wide span of $r_Q$. This gives the value of the DM self-interactions as well as its coupling strength to the Higgs. The value of $r_Q$ is constrained by collider production of the hidden color fermions which is taken up in the following section. 

%%%%%%%%%%%%
\subsubsection{Tuning}
The Higgs potential obtained above also allows us to determine tuning of the Higgs mass parameter. We use the formula
\beq
\Delta=\left| \frac{2\delta\mu^2}{m_h^2}\right|^{-1},
\eeq 
where $\delta\mu^2$ is the leading one-loop correction to the Higgs mass parameter
\beq
\delta\mu^2=\frac{N_c}{8\pi^2}\lambda_t^2\left( m_Q^2-\lambda_t^2f^2\right)\ln\frac{\Lambda^2_\text{UV}}{m_Q^2}=-\frac{3\lambda_t^2\delta_m}{8\pi^2}m_Q^2\ln\frac{\Lambda^2_\text{UV}}{m_Q^2}.
\eeq
Clearly, this tuning depends sensitively on $\delta_m$, and is greatly reduced when $\delta_m\ll 1$.

It is useful to connect $\delta_m$ to $v/f$. This is done by simply minimizing the part of the Higgs potential that depends on $\ln\Lambda_\text{UV}$. This leads to the relation
\beq
\frac{v^2}{f^2}=1-\frac{m_Q^2}{\lambda_t^2f^2}=\frac{\delta_m}{1+\delta_m},
\eeq
similar to what was found in~\cite{Xu:2018ofw}. We rewrite this as
\beq
\delta_m=\frac{v^2/f^2}{1-v^2/f^2},	\label{eq:deltam}
\eeq
to see that $\delta_m$ roughly tracks the tuning required to misalign the vacuum, as it should, for it is by choosing $\delta_m$ small that we obtain the correct Higgs mass. This makes clear that taking $\delta_m$ small is not an additional tuning, but the only tuning required to realize the correct Higgs potential. For instance, when $f\!/\!v=3$ (10) we find $\delta_m\approx0.125$ (0.01) which corresponds to $\sim$10\% (1\%) tuning.

%%%%%%%%%%%%%%%%%%%%%%%%%%%%%%%%%%%
\section{Collider phenomenology}
\label{sec:pheno}
%%%%%%%%%%%%%%%%%%%%%%%%%%%%%%%%%%%
The collider signals of this model arise from the Higgs sector and the production and decay of the hidden color quirks. To determine both these effects we need the physical mass states of the hidden sector fermions. The relevant mass matrix $\mathcal{M}_F$ is
\beq
-\left( \widehat{t}_L, \;\widehat{T}_L\right)\mathcal{M}_F\left(\begin{array}{c}
\widehat{t}_R\\
\widehat{T}_R
\end{array}\right)=-\left( \widehat{t}_L, \;\widehat{T}_L\right)\left(\begin{array}{cc}
m_Q& m_t\\
0 & -c_v\lambda_t f 
\end{array}\right)\left(\begin{array}{c}
\widehat{t}_R\\
\widehat{T}_R
\end{array}\right).
\eeq
As noted in the previous section to obtain the correct Higgs mass without introducing additional fine-tuning, we require, 
\beq
m_Q=\frac{\lambda_t f}{\sqrt{1+\delta_m}}=c_v\lambda_t f, 	\label{eq:mQ}
\eeq
where $\delta_m$ is given in \eq{eq:deltam}. 
In the following, we fix the vector-like mass for the quirk doublet $m_Q$ to the this value. 
Note that we can use this relation to define $f\!/\!v$ in terms of $m_Q$:
\beq
\frac{f}{v}=\sqrt{1+\frac{m_Q^2}{2m_t^2}}.
\eeq
The physical masses are obtained by diagonalizing the fermion mass matrix by the transformation $R(\theta_L)^T\mathcal{M}_FR(\theta_R)$, where the rotation matrices are
\beq
R(\theta_i)=\left(\begin{array}{cc}
\cos\theta_i & \sin\theta_i\\
\!\!\!-\sin\theta_i & \cos\theta_i
\end{array}\right) .
\eeq
The mass eigenvalues are given by
\beq
m^2_\pm=m_Q^2 \left(1+\frac{m_t^2}{2m_Q^2}\pm\frac{m_t}{m_Q}\sqrt{1+\frac{m_t^2}{4m_Q^2}}\right),
\eeq
and the mixing angles are
\beq
\sin2\theta_{L}=-\sin2\theta_R=\frac{1}{\sqrt{1+\frac{m_t^2}{4m_Q^2}}}~.
\eeq
In other words, $\theta_L=-\theta_R\equiv\theta$. The unmixed states are described by Dirac fermions $\widehat{T}_{\pm}$ with masses $m_\pm$, their couplings to SM fields are given in Appendix~\ref{app:FeyQuirk}. 

%%%%%%%%%%%%%%%%%%
\subsection{Scalar Sector}
Like other pNGB Higgs models we find the tree level couplings of the Higgs to SM states are reduced. In our case they are reduced by $c_v$, which follows immediately from Eq.~\eqref{e.csvoverf}. This leads to the usual bound of $f\gtrsim 3v$ from the LHC measurement of Higgs couplings to gauge bosons. It may also lead to interesting signals at the HL-LHC and future colliders. At the same time, the existence of new fermionic states with electric charge that couple to the Higgs amplifies its coupling to photons. As in the quirky little Higgs model~\cite{Cai:2008au}, this pushes the rate of $h\to\gamma\gamma$ closer to the SM prediction~\cite{Burdman:2014zta}. 

\begin{figure}[t]
\centering
\includegraphics[width=0.8\textwidth]{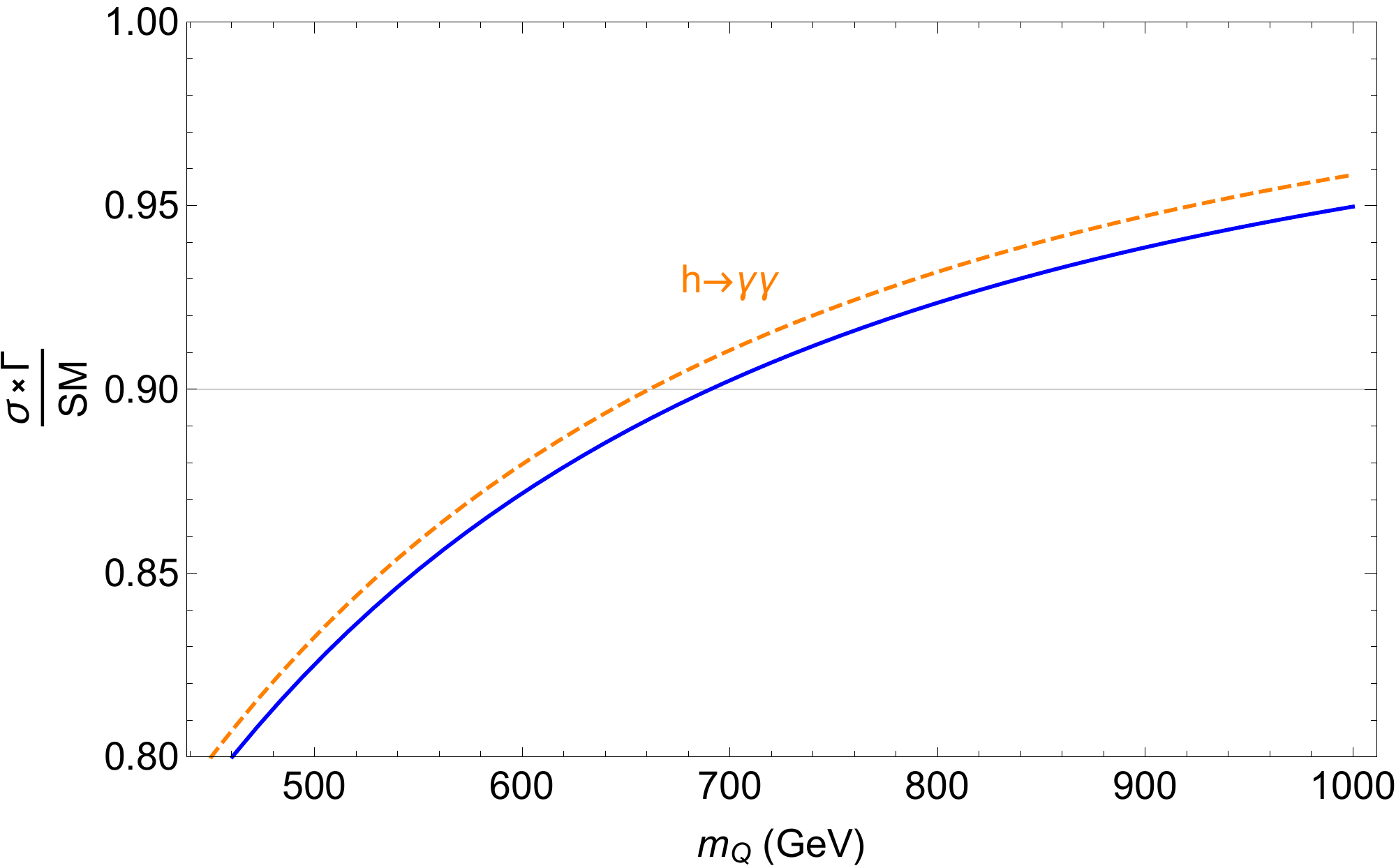} 
\caption{Ratio of the production of a the Higgs boson and subsequent prompt decay into SM final states as a function of $m_Q$. The $h\to\gamma\gamma$ line is given by the dashed orange curve, while all other prompt SM final states fall along the solid blue curve.}
\label{fig:Hcouplings}
\end{figure}

Explicitly, we find the Higgs width into diphotons is approximately
\begin{align}
\Gamma(h\to \gamma\gamma)=\frac{\alpha^2m_h^3}{256\pi^3v^2}&\left|c_v\left[A_V\left(\frac{4m_W^2}{m_h^2} \right)+\frac43A_F\left( \frac{4m_t^2}{m_h^2}\right)\right]\right.\nonumber\\
&+\frac{m_t\sin\theta}{m_{+}}\left(c_v\cos\theta-\sqrt{2}s_v\sin\theta \right)\frac43A_F\left( \frac{4m_{+}^2}{m_h^2}\right)\nonumber\\
&\left. +\frac{m_t\cos\theta}{m_{-}}\left(c_v\sin\theta+\sqrt{2}s_v\cos\theta \right)\frac43A_F\left( \frac{4m_{-}^2}{m_h^2}\right)\right|^2.
\end{align}
In Fig.~\ref{fig:Hcouplings} we see how the production of a given Higgs to SM final state rate changes relative to the SM prediction as a function of $m_Q$. The blue curve shows the usual result for tree level Higgs coupling deviations, while the dashed orange curve denotes the decay into two photons. We see that the latter is slightly increased relative to the other rates. However, the deviation is small enough that it would likely require a future lepton collider to measure it~\cite{Fujii:2017vwa,deBlas:2018mhx,Abada:2019lih}. Current Higgs coupling measurements require this ratio be no less than $0.8$, and the HL-LHC is expected to reach a precision corresponding to about $0.9$~\cite{Cepeda:2019klc}. We see that these already begin to probe $v/f$, but do not reach beyond about 10\% tuning.

The Higgs also develops a loop level coupling to the gluons of the hidden QCD. Similar to coupling to the photon, we find the Higgs coupling to hidden gluons takes the form
\beq
c_{\what g}\frac{\widehat{\alpha}_s}{12\pi}\frac{h}{v}\widehat{G}^a_{\mu\nu}\widehat{G}^{a\mu\nu},
\eeq
where $\widehat{\alpha}_s=\widehat{g}^2_s/(4\pi)$ is the hidden sector strong coupling parameter, $\widehat{G}^a_{\mu\nu}$ is the hidden gluon field strength, and 
\begin{align}
c_{\what g}&=\frac{m_t\sin\theta}{m_{+}}\left(c_v\cos\theta-\sqrt{2}s_v\sin\theta \right)\frac34A_F\left( \frac{4m_{+}^2}{m_h^2}\right)\nonumber\\
&\quad +\frac{m_t\cos\theta}{m_{-}}\left(c_v\sin\theta+\sqrt{2}s_v\cos\theta \right)\frac34A_F\left( \frac{4m_{-}^2}{m_h^2}\right).
\end{align}
This leads to the Higgs width into hidden gluons
\beq
\Gamma(h\to\widehat{g}\widehat{g})=\frac{\widehat{\alpha}_sm_h^5}{288\pi^3v^2}|c_{\what g}|^2,
\eeq
 which may contribute to a detectable Higgs width at future lepton colliders. 

Since the states charged with hidden color carry $U(1)_X$ charge, they are electrically charged under the SM. Bounds from LEP imply that such states cannot be lighter than about 100 GeV. Consequently, the lightest hadrons of the hidden confining group are the glueballs. The lightest glueball state is a $0^{++}$ and has a small mixing with the Higgs. This allows the glueballs to decay with long lifetime to SM states. From~\cite{Juknevich:2009gg} we find the glueball partial width into SM states to be
\beq
\Gamma\left(0^{++}\to X_\text{SM}X_\text{SM}\right)=|c_{\what g}|^2\left[ \frac{\widehat{\alpha}_s}{6\pi v}\frac{f_{0^{++}}}{m_h^2-m_0^2}\right]^2\Gamma_\text{SM}\left(h(m_0)\to X_\text{SM}X_\text{SM}\right)\label{e.Gdecay}
\eeq
 where $m_0$ is the mass of the lightest glueball, $f_{0^{++}}=\langle 0|\text{Tr}\,\widehat{G}_{\mu\nu}\widehat{G}^{\mu\nu}|0^{++}\rangle$, and $\Gamma\left(h(m_0)\to X_\text{SM}X_\text{SM}\right)$ is the SM Higgs partial width for a Higgs with mass $m_0$. Lattice calculations have determined $4\pi \widehat{\alpha}_sf_{0^{++}}=3.1m_0^3$~\cite{Chen:2005mg}. In addition, the exotic decays of the Higgs into glueballs with displaced decays can lead to striking signals at the LHC~\cite{Curtin:2015fna}.

\begin{figure} [t]
\centering
\includegraphics[width=0.8\textwidth]{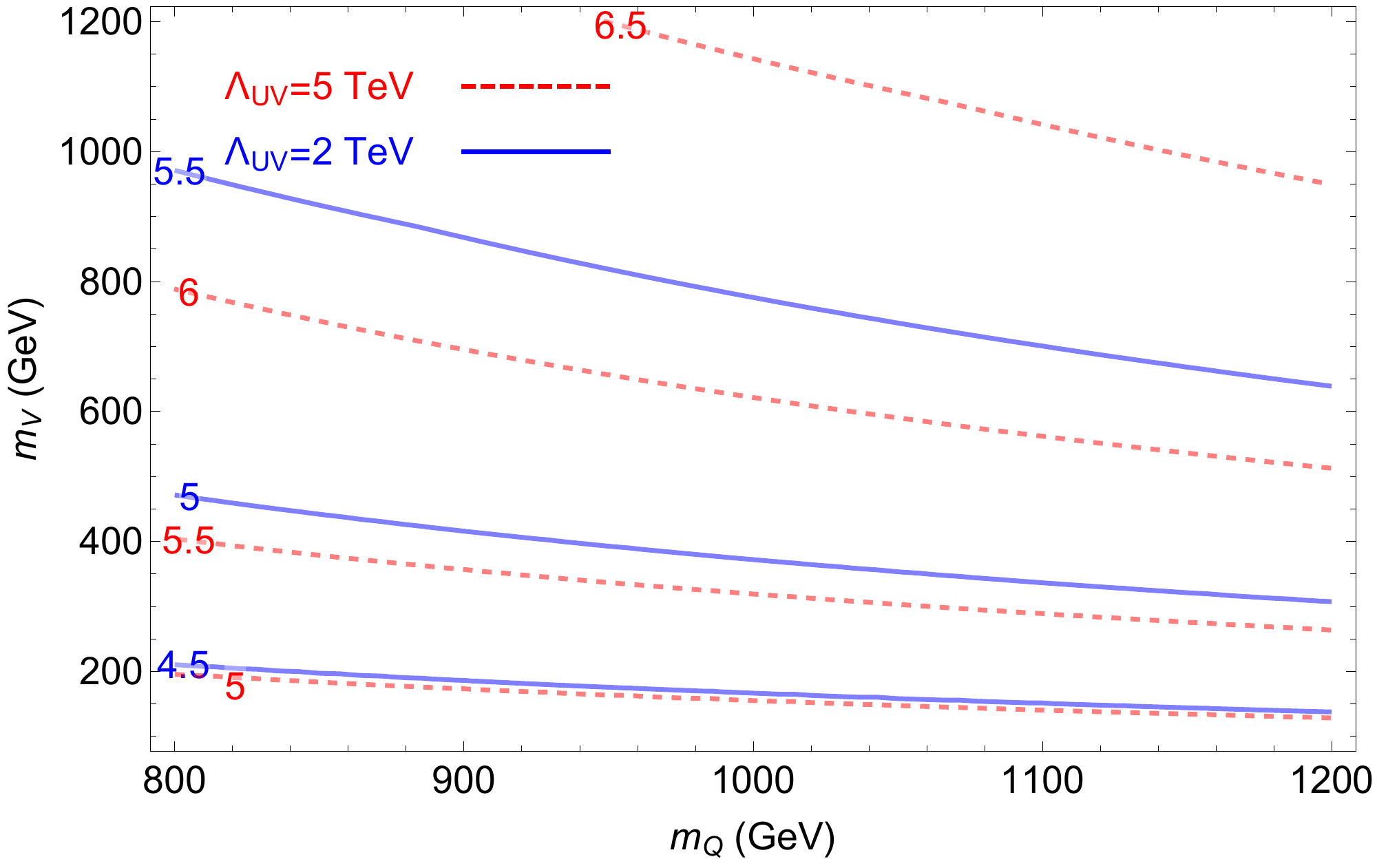} 
\caption{Contours of the hidden color $\widehat{\Lambda}_\text{QCD}$ as a function of the vector-like masses of hidden color fermions. The blue solid (red dashed) contours correspond to a UV cutoff $\Lambda_{\rm UV}\!=\!2 (5) \tev$ where the SM and hidden strong couplings become equal.}
\label{fig:lamQCD}
\end{figure}

To be more precise we must estimate the mass of the hidden glueball. This is done by estimating the hidden scale $\widehat{\Lambda}_\text{QCD}$ using two-loop running.\footnote{While this scale has its drawbacks~\cite{DeGrand:2019vbx} in the pure gauge limit there are not many physical scales to choose from.} We assume at scales near the cutoff of a few TeV the SM and hidden strong couplings become equal because of the $\mathbb{Z}_2$ symmetry in the UV. Thus, we can run the SM strong coupling up to the cutoff and then run the hidden coupling down from the cutoff for a given spectrum. In Fig.~\ref{fig:lamQCD} we find that the hidden color strong scale varies between about 4.5 to 6.5 GeV for $m_Q\in[800,1200]\gev$. This implies the lightest glueball mass, taken to be about $6.8\widehat{\Lambda}_\text{QCD}$, is likely to fall between 30 and 45 GeV. Then using the glueball decay width in Eq.~\eqref{e.Gdecay} we find the glueballs typically have a decay length of hundreds of meters, with the smallest values for larger $m_V$ and $\Lambda_\text{UV}$. The displaced decays from these particles may be quite challenging for the ATLAS and CMS to detect, but may be detected by MATHUSLA-like detectors~\cite{Curtin:2018mvb}.

There may also be new scalars related to the spontaneous symmetry breaking mechanism. In weakly coupled UV completions there may be a radial mode, a scalar whose mass close to $f$. As has been detailed for other pNGB realizations of neutral naturalness~\cite{Craig:2015pha,Buttazzo:2015bka,Ahmed:2017psb,Chacko:2017xpd,Kilic:2018sew,Alipour-fard:2018mre}, this scalar will have order one couplings to all the pNGBs, leading to observable signals at the LHC and future colliders. If the UV completion involves an approximate scale symmetry then a heavy scalar associated with the breaking of scale invariance, the dilaton, can have large coupling to the SM and hidden sector states~\cite{Ahmed:2019kgl} providing additional collider signals. 

\subsection{Electroweak Precision Tests}
\label{sec:EWPT}
%%%%%%%%%%%%%%%%%%%%%%%%%%%%%%%%%%%
Extensions of the SM are constrained by precision electroweak measurements. The constraints can be expressed in terms of the oblique parameters $S$, $T$, and $U$~\cite{ Peskin:1990zt,Peskin:1991sw}. The contributions to $U$ are typically small, so we only compute the contribution to $S$ and $T$. These contributions arise from the new electroweak charged fermions inducing important radiative corrections to gauge boson propagators. In addition, the modified coupling of the Higgs boson to gauge bosons leads to an infrared log divergence~\cite{ Contino:2010rs}. We find the leading contributions to be
\begin{align}
S\approx&\frac{2N_{\widehat c}\,m_t^2}{15\pi  \,m_Q^2}+\frac{1}{12\pi}\frac{v^2}{f^2}\ln{\left(\frac{\Lambda^2_{\rm UV}}{m_h^2}\right)}
   ,\label{e.S}\\
 T\approx&\frac{13N_{\widehat c}\, m_t^4} {120 \pi  m_Z^2 m^2_Q\sin^22\theta_W }-\frac{3}{16\pi}\frac{1}{\cos^2
   \theta _W}\frac{v^2}{f^2}\ln{\left(\frac{\Lambda_{\rm UV}^2}{m_h^2}\right)},\label{e.T}
     \end{align}
where $\Lambda_{\rm UV}$ is UV cutoff scale, $\theta_W$ is the usual weak mixing angle, and the factor of $N_{\widehat c}$ comes from the number of dark QCD color. 

\begin{figure} [t!]
\centering
\includegraphics[width=0.67\textwidth]{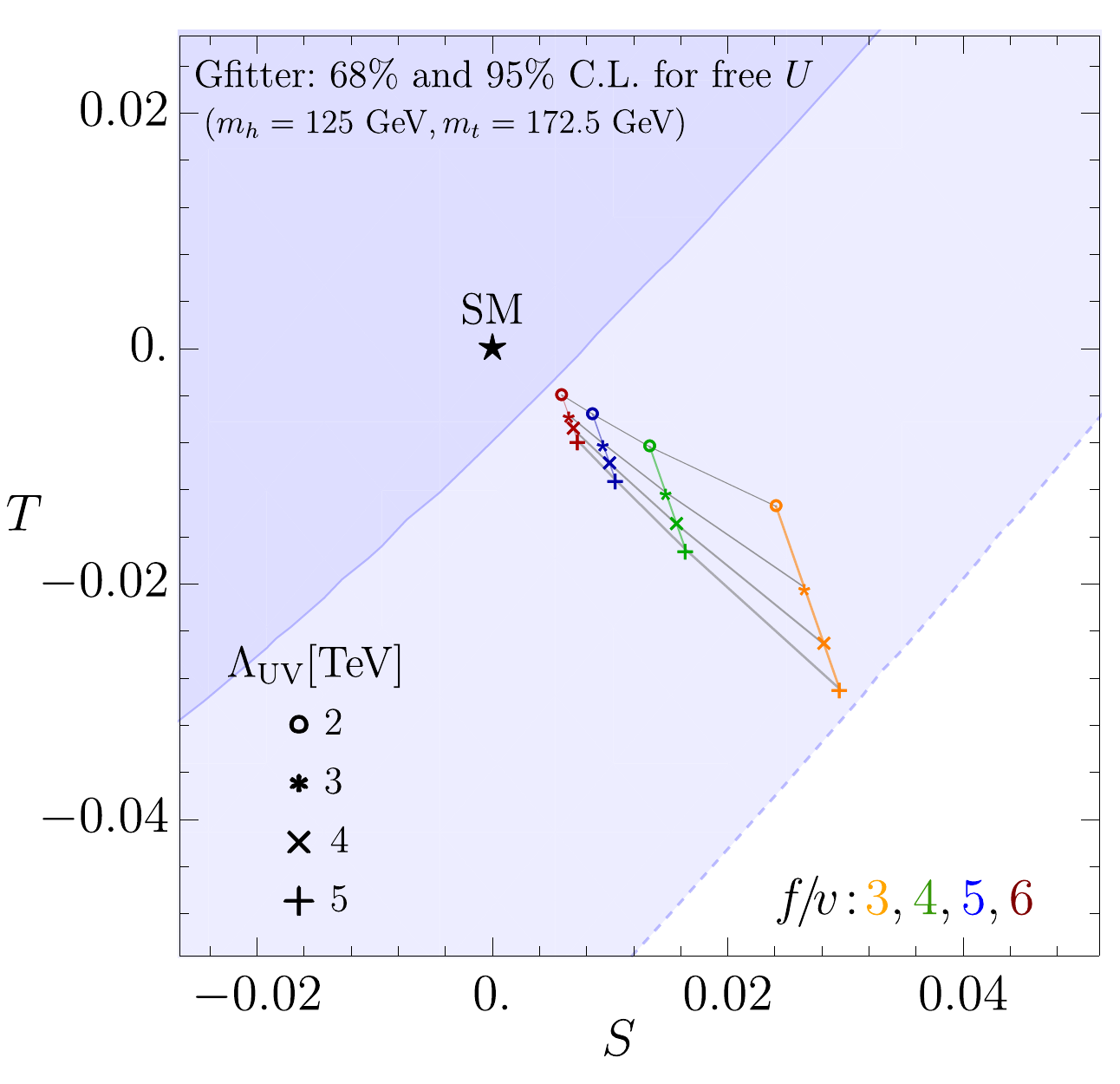} 
\caption{The allowed region in the $S\!-\!T$ plane leaving the $U$ parameter free \cite{Haller:2018nnx}. The colored points (orange, green, blue, and red) indicate the values of $S$ and $T$ for $f\!/\!v=3,4,5,6$ with UV cutoff scale $2,3,4, 5$ TeV.}.
\label{fig:ST}
\end{figure}
As expected, the contributions from vector-like fermions, $\widehat{X}$ and $\widehat{Y}$, cancel as well as the power law UV divergences. 
These contributions are compared to the experimental fits and found to lie within the $68\%$ and $95\%$ allowed regions as provided by the Gfitter collaboration~\cite{Haller:2018nnx}. 
In Fig.~\ref{fig:ST}, we plot $S$ and $T$ with $U$ free, for the input parameters $m_h=125$ GeV and $m_t=172.5$ GeV. The colored points in the figure correspond to values of $f\!/\!v=3, 4, 5, 6$ (orange, green, blue, and red) and $\Lambda_\text{UV}$, 2, 3, 4, 5 TeV. With increasing the value of $f$, the value of $S$ and $T$ approach the SM value.
%

%%%%%%%%%%%%%%%%%%%%%%%%%%%%%%%%%%%%%%%%
\subsection{Quirky Signals \label{s.QuirkySignals}}
\begin{figure} [t]
\centering
\includegraphics[width=0.8\textwidth]{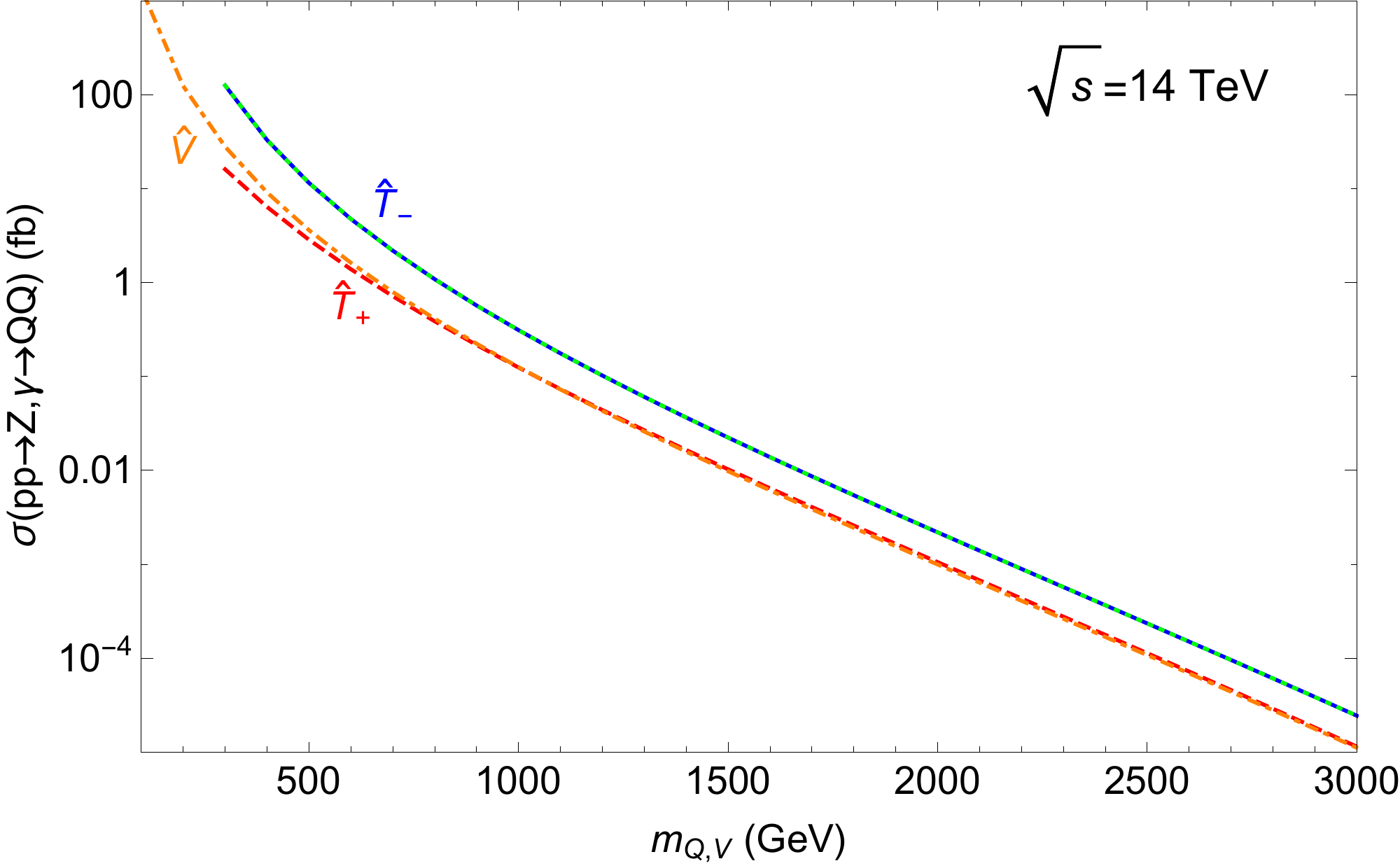} 
\caption{Production cross sections for quirks as a function of the relevant vector-like mass at a 14 TeV proton collider for $\delta_m=0.1$. The production of $\widehat{T}_+\widehat{T}_-$ is given by the dotted green line, which lies nearly on top of the blue $\widehat{T}_-\widehat{T}_-$ curve. }
\label{fig:qirkProd}
\end{figure}

The new fermions ($\widehat{T}_{\pm}$, $\widehat{X}$, and $\widehat{Y}$) can be produced at colliders through Drell-Yan due to their hypercharge of 2/3. We parameterize the couplings of any fermion $f$ to the $Z$ boson and the photon by 
\beq
\mathcal{L}\supset\frac{g}{2c_W}Z_\mu\overline{f}\gamma^\mu(v_f-a_f\gamma^5)f+eQ_fA_\mu\overline{f}\gamma^\mu f,
\eeq
where $c_W$ and $s_W$ is the cosine and sine of the weak mixing angle while $g$ and $e=g s_W$ are the weak and electric couplings, respectively. As an example, SM fermions have $v_f=T_3-2Qs_W^2$ and $a_f=T_3$. We then find the partonic cross section for $\overline{q}q\to Z,\gamma\to \overline{f}f$ to be
\begin{align}
\sigma(\overline{q}q\to \overline{f}f)(\tilde{s})=&\frac{\pi\alpha_Z^2N_{\what{c}}}{12 N_c \tilde{s}(1-m_Z^2/\tilde{s})^2}\sqrt{1-\frac{4m_f^2}{\tilde{s}}}\nonumber\\
&\times\left\{\Big(1+\frac{2m_f^2}{\tilde{s}} \Big)\left[\left|v_qv_f+4s_Wc_WQ_qQ_f\left(1-\tfrac{m_Z^2}{\tilde{s}} \right) \right|^2+\left| a_qv_f\right|^2 \right] \right.\nonumber\\
&\left.\phantom{AA}+\Big(1-\frac{4m_f^2}{\tilde{s}} \Big)\left[\left|v_qa_f \right|^2 +\left|a_qa_f \right|^2 \right]\right\},
\end{align}
where $\alpha_Z\equiv g^2/(4\pi c_W^2)$. In Fig.~\ref{fig:qirkProd} we see the fermion cross sections at a 14 TeV proton-proton collider. We used MSTW2008 PDFs~\cite{Martin:2009iq}  and a factorization scale of $\sqrt{\tilde{s}}/2$.

All the fermions charged under the hidden color group have masses much above 100 GeV due to LEP bounds on charged particles. The hidden confining scale is of the order of a few GeV, so we expect them to exhibit quirky~\cite{Kang:2008ea} dynamics, which can give a variety of new signals at colliders~\cite{Harnik:2011mv,Farina:2017cts,Knapen:2017kly,Evans:2018jmd,Li:2019wce,Li:2020aoq}. After production they are connected by a string of hidden color flux which, because there are no light hidden color states, is stable against fragmentation. The quirky pair behaves as though connected by a string with tension $\sigma\sim3.6\widehat{\Lambda}_\text{QCD}^2$~\cite{Lucini:2004my}, see also~\cite{Teper:2009uf}. 

Much of the subsequent dynamics can be treated semi-classically. Since these quirks carry electric charge the oscillating particles radiate soft-photons, quickly shedding energy until they reach their ground state~\cite{Burdman:2008ek,Harnik:2008ax}. Annihilation is strongly suppressed in states with nonzero orbital angular momentum, so in nearly every case the quirks do not annihilate until they reach the ground state. Since the quirks are accelerated by the string tension, we can estimate their acceleration as $a=\sigma/m_f$. Then, using the Larmor formula we can estimate the radiated power as
\beq
\mathcal{P}=\frac{8\pi \alpha}{3}a^2=\frac{8\pi \alpha\sigma^2}{3m_f^2},
\eeq
where $\alpha=e^2/(4\pi)$. The time it takes the quirky bound state to drop to its ground state is given by $K/\mathcal{P}$, where $K$ is the kinetic energy of the quirks. Taking $K\sim m_f$ we can then estimate the de-excitation time $T_d$ as
\beq
T_d\sim\frac{3m_f^3}{8\pi \alpha \left( 3.6\widehat{\Lambda}_\text{QCD}^2\right)^2}\sim 4\times10^{-19}\,\text{sec}\left(\frac{m_f}{\text{800 GeV}} \right)^3\left(\frac{\text{6 GeV}}{\widehat{\Lambda}_\text{QCD}} \right)^4.
\eeq
Clearly, the de-excitation is very fast, leading to prompt annihilation

Depending on the masses of the hidden $\widehat{b}$ quark, the $\widehat{T}_\pm$ could $\beta$-decay by emitting a $W$. When the mass splitting is small the de-excitation is faster and the states typically annihilate. However, if the splitting is large it is most likely that both top-like states transition to bottom-like states. These would then de-excite and annihilate in the same way, though there would be additional $W$s in the final state.

If the $\widehat{b}$ quarks are not too heavy, then $\widehat{T}_\pm\widehat{b}$ combinations can be produced through the $W$ boson. If these states are similar in mass so that $\beta$-decay is slow then the bound states can lead to visible signals, like $W\gamma$ resonances, with appreciable rates. This is because the electric charge of the state prevents its decay into hidden gluons. However, larger splittings allow the heavier state to decay to the lighter promptly, diluting these signals significantly.

Because the quirks are fermions there are four $s$-wave states, one singlet and three triplet. Following~\cite{Cheng:2018gvu} we assume that each of these states is populated equally by production, so we take the total width $\Gamma_{\rm tot}$ of the bounds state to be 
\beq
\Gamma_{\rm tot}=\Gamma_s+3\Gamma_t,
\eeq
where $\Gamma_s$ and $\Gamma_t$ are the widths of the singlet and triplet states respectively. 

\begin{figure} [t!]
\centering
\includegraphics[width=0.49\textwidth]{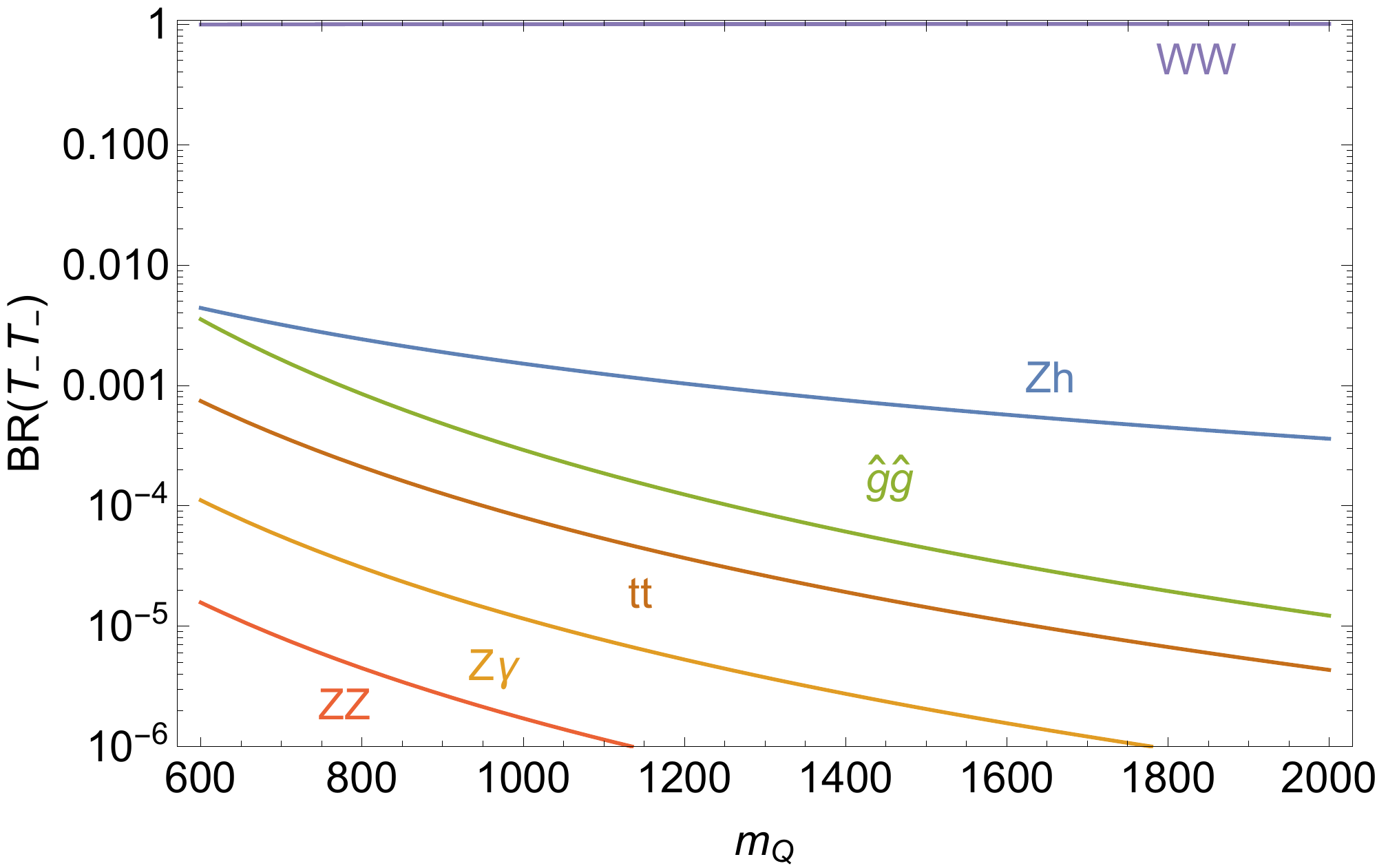} 
\includegraphics[width=0.49\textwidth]{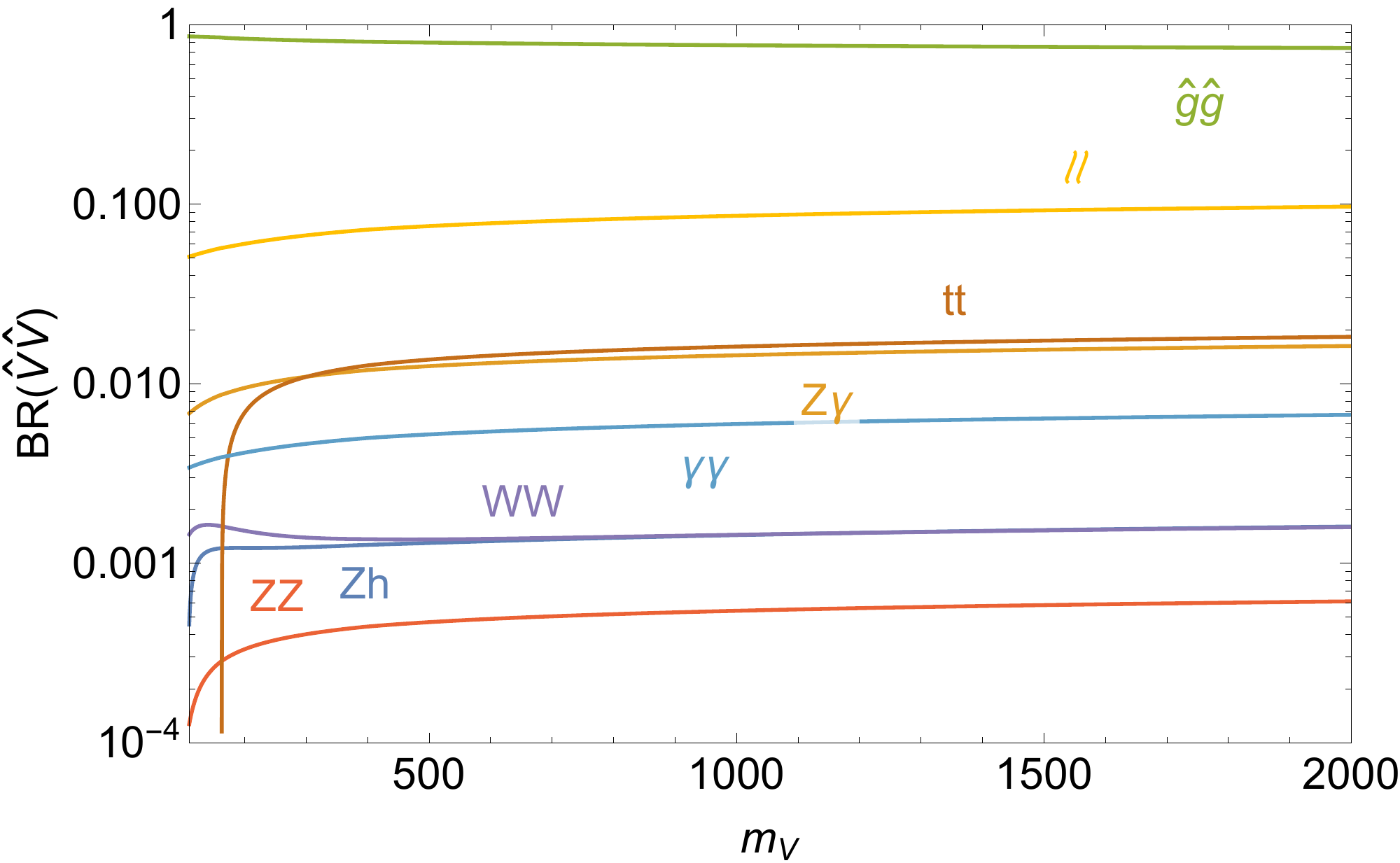} 
\caption{Branching fraction for the $s$-wave quirkonium composed of the $\widehat{T}_-$ (left) and $\widehat{X}$ or $\widehat{Y}$ (right) quirks. The $\widehat{T}_-$ quirks dominant decays are into weak gauge bosons $W^+W^-$. The dominant branching fraction for the $\widehat{X}$ and $\widehat{Y}$ is into hidden gluons.  }
\label{fig:qirkDecay}
\end{figure}

\begin{figure} [t!]
\centering
\includegraphics[width=0.49\textwidth]{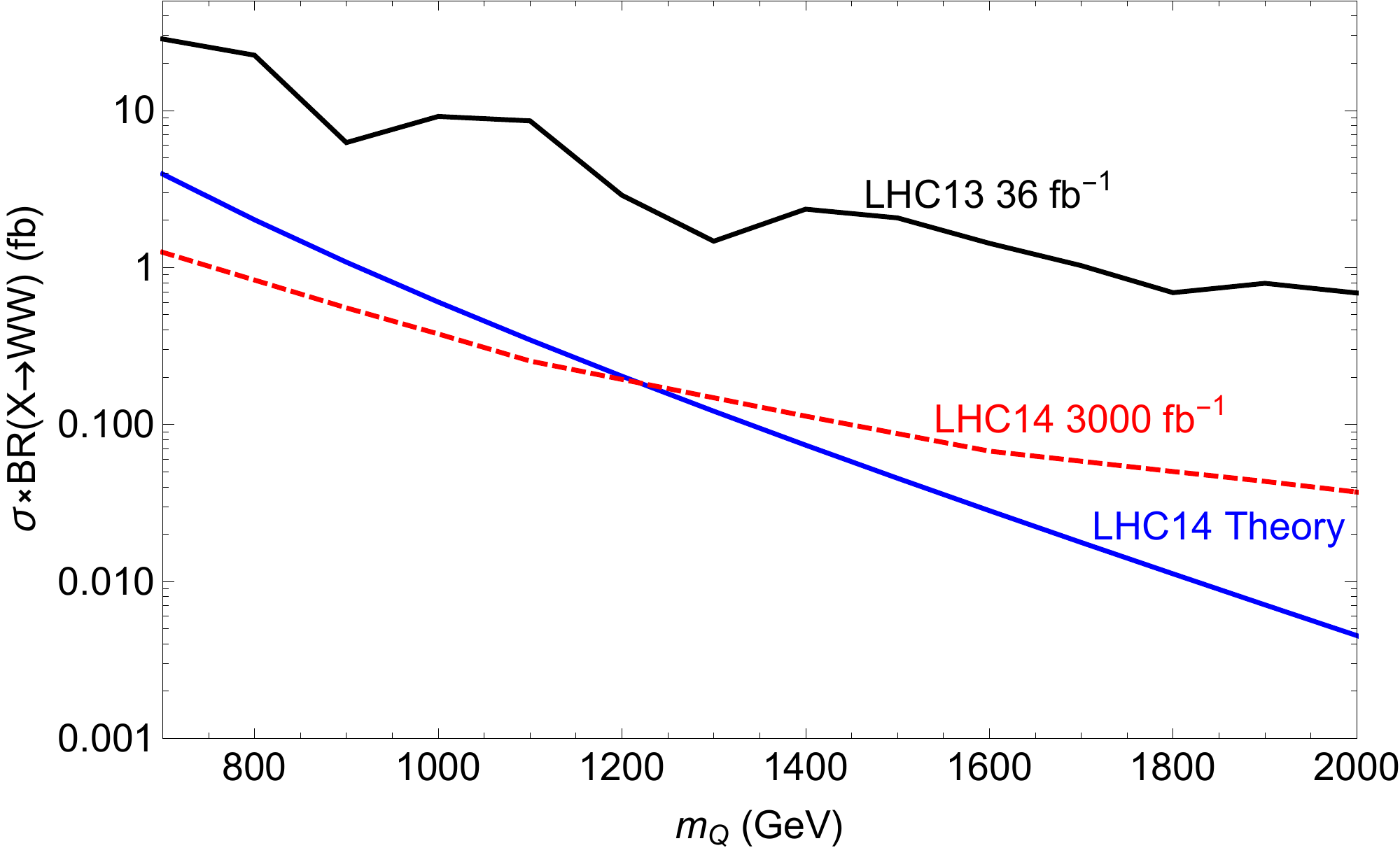} 
\includegraphics[width=0.49\textwidth]{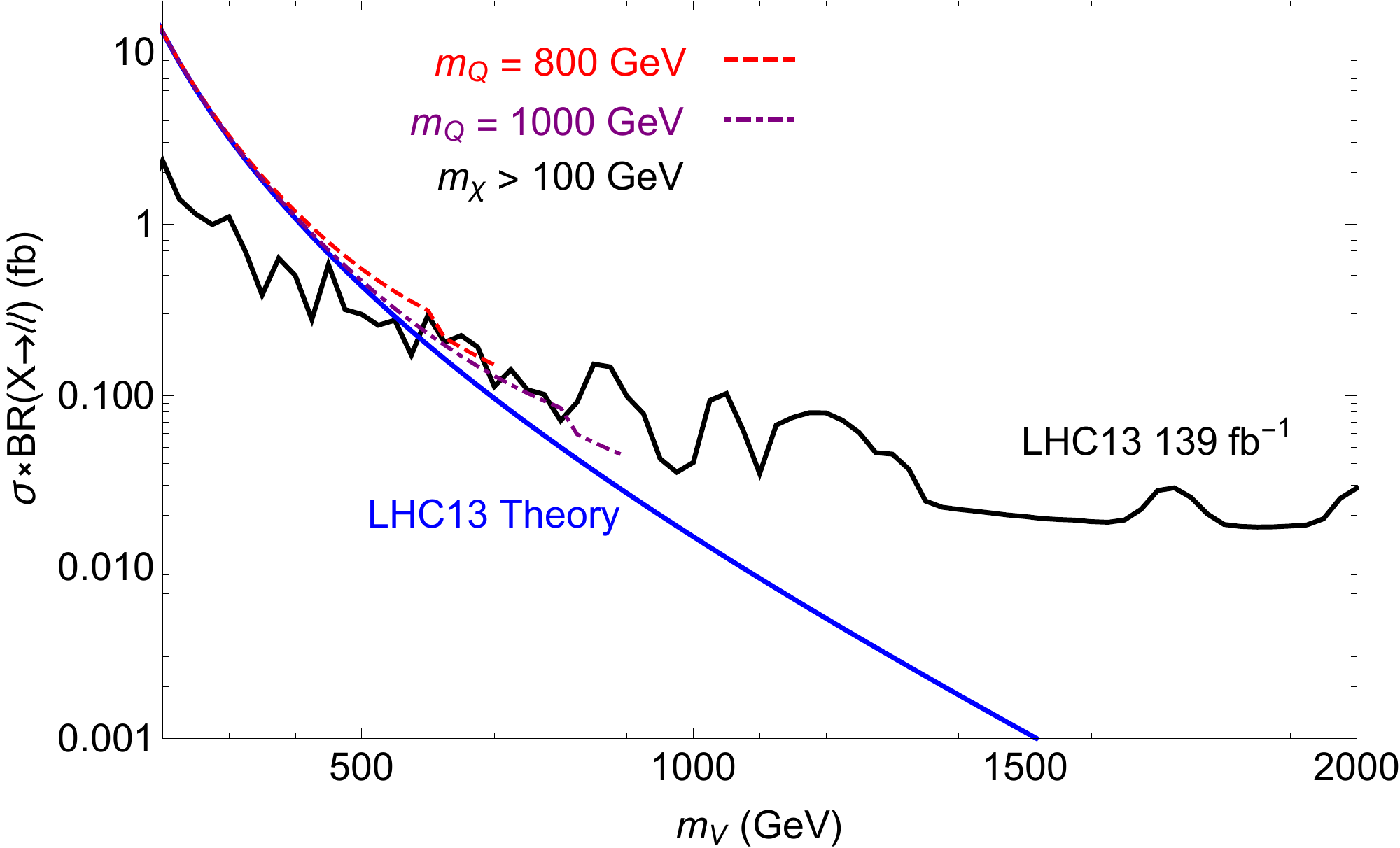} 
\caption{{\it Left:} Comparison of LHC 13 reach in $W^+W^-$ resonance searches for a $\widehat{T}_{-}\widehat{T}_{-}$ bound state (black) to the theoretical 14 TeV prediction (blue) and expected sensitivity of the high luminosity run (dashed red). {\it Right:} Comparison of LHC 13 reach in dilepton resonance searches for $\widehat{X}\widehat{X}$ and $\widehat{Y}\widehat{Y}$ bound states (black) to the theoretical 13 TeV prediction (blue). Contributions from $\widehat{T}$ states included for $m_Q=800$ (1000) GeV in red dashed (purple dash-dotted) curve for DM masses greater than 100 GeV.}
\label{f.WWsearch}
\end{figure}

For the $\widehat{T}_\pm\widehat{T}_\pm$ states which carry weak isospin the dominant quirkonium decays are to $WW$ with a branching fraction of about 75\%. This comes from the chiral enhancement in this decay. This signal has been searched for at the LHC by both ATLAS~\cite{Aaboud:2017gsl,Aaboud:2017fgj} and CMS~\cite{Sirunyan:2016cao,Sirunyan:2017acf}. The next largest fractions are into $Zh$, at the 10\% level, which can be compared to ATLAS~\cite{Aaboud:2017cxo} and CMS~\cite{Khachatryan:2016cfx,Sirunyan:2018fuh} searches. All other visible final states are suppressed well below the percent level, see Fig.~\ref{fig:qirkDecay}. Of these, the most likely LHC signal is a new scalar resonance decaying to $WW$, though this does depend on the $\widehat{b}$-quirk mass. As shown in Fig.~\ref{f.WWsearch}, current searches are not yet sensitive to these signals. Here we assume a production of the $\widehat{T}_-\widehat{T}_-$ directly, and through production of the $\widehat{T}_+$ state which then decays to a soft $Z$ and $\widehat{T}_-$. While the LHC is not yet sensitive to these signals, the high luminosity run (dashed red line) will probe the most natural regions of parameter space~\cite{ATL-PHYS-PUB-2018-022}.

The $\widehat{X}$ and $\widehat{Y}$ particles only couple to visible states through hypercharge, hence there is no rate into $Zh$ and the rate into $WW$ vanishes when the $Z$ mass can be neglected. The largest coupling is to hidden gluons, so this dominates the branching fractions. These gluons shower and hadronize into hidden QCD glueballs, some fraction which may have displaced decays at the LHC~\cite{Chacko:2015fbc}. However, they can also annihilate into $\bar{f}f$ and EW gauge bosons through their hypercharge coupling, see Fig.~\ref{fig:qirkDecay}. Of these, dilepton and diphoton channels have the greatest discovery potential because the signal is so clean, which has motivated searches at both ATLAS~\cite{Aaboud:2017yyg,Aad:2019fac} and CMS~\cite{Sirunyan:2018wnk,CMS:2019tbu}. In the right panel of Fig.~\ref{f.WWsearch} we compare the reach of the ATLAS search~\cite{Aad:2019fac} to the theoretical prediction. We see that quirks below about 550 GeV are in tension with current collider bounds. Seeing that the predicted cross section is near the experimental limit, it is likely that by the end of the LHC run 3, with 300 $\text{fb}^{-1}$, any quirks of this type below a TeV will be discovered. Further LHC runs can probe even larger $m_{\widehat{V}}$, but we note that taking this mass larger does not affect the naturalness of the Higgs mass. It does, however, indicate that the DM is heavier, see Eq.~\eqref{e.chiMass}. 

When $m_V>m_-+m_\chi$ the $\widehat{X},\widehat{Y}$ quirks will quickly decay, $\widehat{V}\to \widehat{T}_-+\chi$. In this case the powerful dilepton resonance search will not apply. Instead, the production cross section for  $T_-$ bound states must include this, in general small, additional mode. A similar story holds if $m_->m_V+m_\chi$, where now the $\widehat{T}_-$ quirk decays promptly to an $\widehat{X}$ or $\widehat{Y}$ and a DM scalar. Then, the dilepton bounds would apply to the $\widehat{T}$ production. For lighter $m_Q$ this can strengthen the bound on $m_V$. The red dashed and purple dash-dotted lines on the dilepton bound in Fig.~\ref{f.WWsearch} correspond to taking $m_Q=800$ GeV and $m_Q=1000$ GeV, respectively, and the DM mass of 100 GeV. By taking the DM heavier these lines would cut off earlier, at $m_V=m_\pm-m_\chi$. 

In summary, standard collider searches for prompt visible objects do constrain $m_{\widehat{V}}>550$ GeV, but the other parameters of the model are less restricted. However, both the displaced searches related to the hidden sector glueballs and dilepton and diboson resonance searches can provide evidence for the hidden QCD sector at the LHC. As we shall see in the next section, this parametric freedom can lead to viable DM, and complementary search strategies from DM experiments. 

%%%%%%%%%%%%%%%%%%%%%%%%%%%%%%%%%%%
\section{Dark matter phenomenology}
\label{sec:DMpheno}
%%%%%%%%%%%%%%%%%%%%%%%%%%%%%%%%%%%
In this section we detail the phenomenology of the DM candidate $\chi$, the complex scalar charge under the global symmetry $U(1)_D$. As mentioned above, this global symmetry stabilizes the DM. 
All the SM fields and the quirky top partners $\widehat{T}_\pm$ are $U(1)_D$ neutral, whereas the quirky fermions $\widehat X$ and $\widehat Y$ are charged. 
The $U(1)_D$ global symmetry is exact, so we can associate a discrete dark $\mathbb{Z}_2$ parity under which, 
\beq
\chi \overset{\mathbb{Z}_2}{\to} -\chi, \qquad \widehat X \overset{\mathbb{Z}_2}{\to} -\widehat X, \qquad \widehat Y \overset{\mathbb{Z}_2}{\to}-\widehat Y,
\eeq
but more generally we simply consider particles in this sector as carrying a global dark charge, which prevents their decay. Since the quirky states $\widehat X$ and $\widehat Y$ have the fractional SM electric charge $2/3$ they cannot be the DM. However, the SM neutral complex scalar $\chi$ is our DM candidate as long as it is the lightest $U(1)_D$ charged particle. 

To determine the success of this scalar as explaining the observed DM in the universe, in what follows we calculate the relic abundance and DM-nucleon cross section for the direct detection in our model. We then consider the dark matter annihilation for the indirect detection and impose the collider constraints on our parameter space. We find that much of the natural parameter space of this model has not yet been conclusively probed by experiment, but is expected to be covered next several years.

%%%%%%%%%%%%%%%%%%%%
\subsection{Relic abundance\label{s.relic}}
\begin{figure}[t!]
\centering
\includegraphics[width=\textwidth]{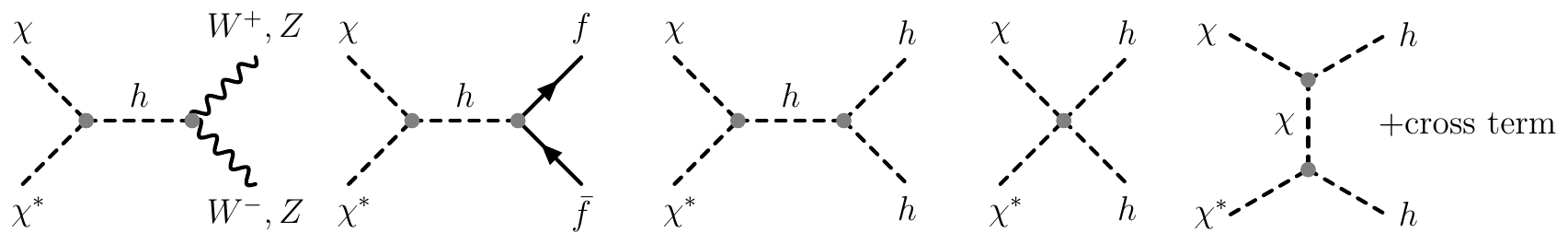} 
\caption{The Feynman diagrams for the DM annihilation to SM.}
\label{fig:chichi_ann}
\end{figure}

\begin{figure}[t!]
\centering
\includegraphics[width=\textwidth]{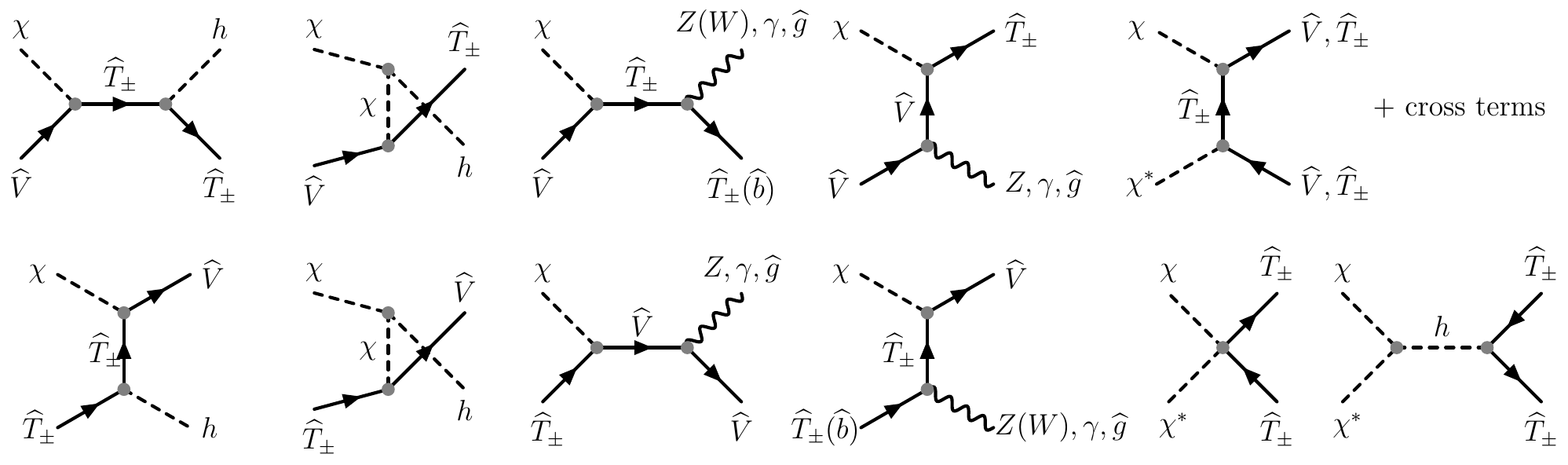}
\caption{The Feynman diagrams for the DM semi-annihilation/conversion through the dark quirks.}
\label{fig:chichi_semiann}
\end{figure}

%%%%%%%%%%%%%%%%%%%%
The thermal relic density of the scalar $\chi$ is obtained using the standard freeze-out analysis. Figures~\ref{fig:chichi_ann} and \ref{fig:chichi_semiann} show the relevant Feynman diagrams for the DM annihilation and semi-annihilation/conversion, respectively. The Boltzmann equation for the DM annihilation and semi-annihilation/conversion processes is
\begin{align}
\frac{dn_{\chi}}{dt}=& -3{\cal H} n_{\chi}-\langle\sigma^{\chi\chi\phi\phi^\p}_{\vmol}\rangle\Big(n_{\chi}^2-\bar n_{\chi}^2\Big)	\notag\\
&-\langle\sigma^{X\what V\what T_{\!\pm} \phi}_{\vmol}\rangle\Big(n_{\chi}n_{\what V}-\bar n_{\chi} \bar n_{\what V}\frac{n_{\what T_{\!\pm}}}{\bar n_{\what T_{\!\pm}}}\Big) -\langle\sigma^{\chi\what T_{\!\pm}\what V \phi}_{\vmol}\rangle\Big(n_{\chi}n_{\what T_{\!\pm}}-\bar n_{\chi} \bar n_{\what T_{\!\pm}}\frac{ n_{\what V}}{\bar n_{\what V}}\Big)	\notag\\
&-\langle\sigma^{\chi\chi\what T_{\!\pm} \what T_{\!\pm}}_{\vmol}\rangle\Big(n_{\chi}^2- \bar n_{\chi}^2\frac{n_{\what T_{\!\pm}}^2}{\bar n_{\what T_{\!\pm}}^2}\Big) -\langle\sigma^{\chi\chi\what V \what V}_{\vmol}\rangle\Big(n_{\chi}^2- \bar n_{\chi}^2\frac{n_{\what V}^2}{\bar n_{\what V}^2}\Big),
\label{eq.boltzmann}
\end{align}
where $\phi (\phi^\prime)$ are the SM fields: $h, t,W,Z,\gamma,\cdots$. Also, ${\cal H}$ is the Hubble parameter and $n_i$ is the number density of species $i$, whereas the $\bar n_i$ is its thermal equilibrium value. The quantity $\langle\sigma^{ij kl}_{\vmol} \rangle\!\equiv\! \langle\sigma(ij\to kl)\,{\vmol}\rangle$ is the thermal averaged cross-section of the initial states $ij$ to final states $kl$ with $\vmol$ being the M{\o}ller velocity.  The last term in the first line of \eq{eq.boltzmann} describes the dynamics of the standard DM annihilation to the SM final states as shown in \fig{fig:chichi_ann}. The second and third lines describe the semi-annihilation and conversion processes shown in \fig{fig:chichi_semiann}.

The dominant DM annihilation channels are to the SM, i.e. $\chi\chi^\ast\to WW,hh,ZZ,t\bar t,b\bar b$, while the semi-annihilation and conversion processes are only relevant if the masses the quirk states ($\what V, \what T_\pm$) are similar to $m_\chi$. When the quirk masses are much larger than the DM, their thermal distributions are Boltzmann suppressed, making  semi-annihilation or conversion processes very rare as compared to the standard annihilation processes. The relevant Feynman rules to calculate the DM annihilation or semi-annihilation processes are given in \app{app:FeyQuirk}. The DM relic abundance is computed using the public code \texttt{micrOMEGAs}~\cite{Belanger:2018mqt}. 

Before discussing these results we emphasis some of the features of this model. 
\begin{itemize}\itemsep0em
\item The top partners are SM color neutral, therefore the symmetry breaking scale $f$ may be at or below a TeV. This leads to significant improvements in the fine-tuning while simultaneously allowing a larger window for the pNGB DM masses in comparison to colored top partner models~\cite{Balkin:2017aep,Frigerio:2012uc,DaRold:2019ccj}.
\item The DM annihilations to SM are dominated by $s$-channel Higgs exchange. The amplitude for such processes is,
\beq
{\cal M}_{\chi\chi \phi\phi^\prime}\propto \Big(\frac{s}{f^2}-2\lambda_{h\chi}\Big)v,
\eeq
where $s\!=\!4 m_\chi^2$. The $s$ dependent term originates from the derivative coupling $\partial_\mu h\,\partial_\mu(\chi^\ast\chi)$, while the $\lambda_{h\chi}$ term is a loop induced explicit breaking of the $\chi$ shift symmetry, see \eq{eq:eff_pot}. 
\item When the standard DM annihilation processes dominate (which we see below is typically the case), the DM relic abundance can be estimated as, 
\beq
\Omega_\chi h^2\approx 0.12 \left(\frac{2.2\!\times \!10^{-26} \,{\rm cm}^3{\rm s}^{-1}}{\langle\sigma(\chi\chi^\ast\to {\rm SM})\vmol\rangle}\right),
\eeq
where $0.12$ is the observed DM relic abundance by the PLANCK satellite~\cite{Aghanim:2018eyx}. 
\item The thermal averaged annihilation cross section to SM fields via $s$-channel Higgs exchange is proportional to 
\beq
\langle\sigma(\chi\chi^\ast\to {\rm SM})\vmol\rangle \propto \frac{1}{m_\chi^2} \Big(\frac{4m_\chi^2}{f^2}-2\lambda_{h\chi}c_v^2\Big)^2,
\eeq
which implies that in the limit $\lambda_{h\chi}\to 0$, i.e. no explicit shift symmetry breaking, the cross section is proportional to $m_\chi^2/f^4$. 
Hence, for a given $m_\chi$ the relic abundance, $\Omega_\chi h^2$, scales as $f^4$. 
\item For $m_\chi^2/f^2\ll 1$, $\chi$ annihilation proceeds through the portal coupling $\lambda_{h\chi}$. 
When $m_\chi^2\sim \lambda_{h\chi} f^2/2$ the annihilation cross-section drops due to cancellation between the $s$-channel process, enhancing the relic abundance. For $m_\chi^2\gg \lambda_{h\chi} f^2/2$ the DM relic abundance falls like $1/m_\chi^2$ for fixed~$f$. 
\end{itemize}

\begin{figure}[t!]
\centering
\includegraphics[width=0.5\textwidth]{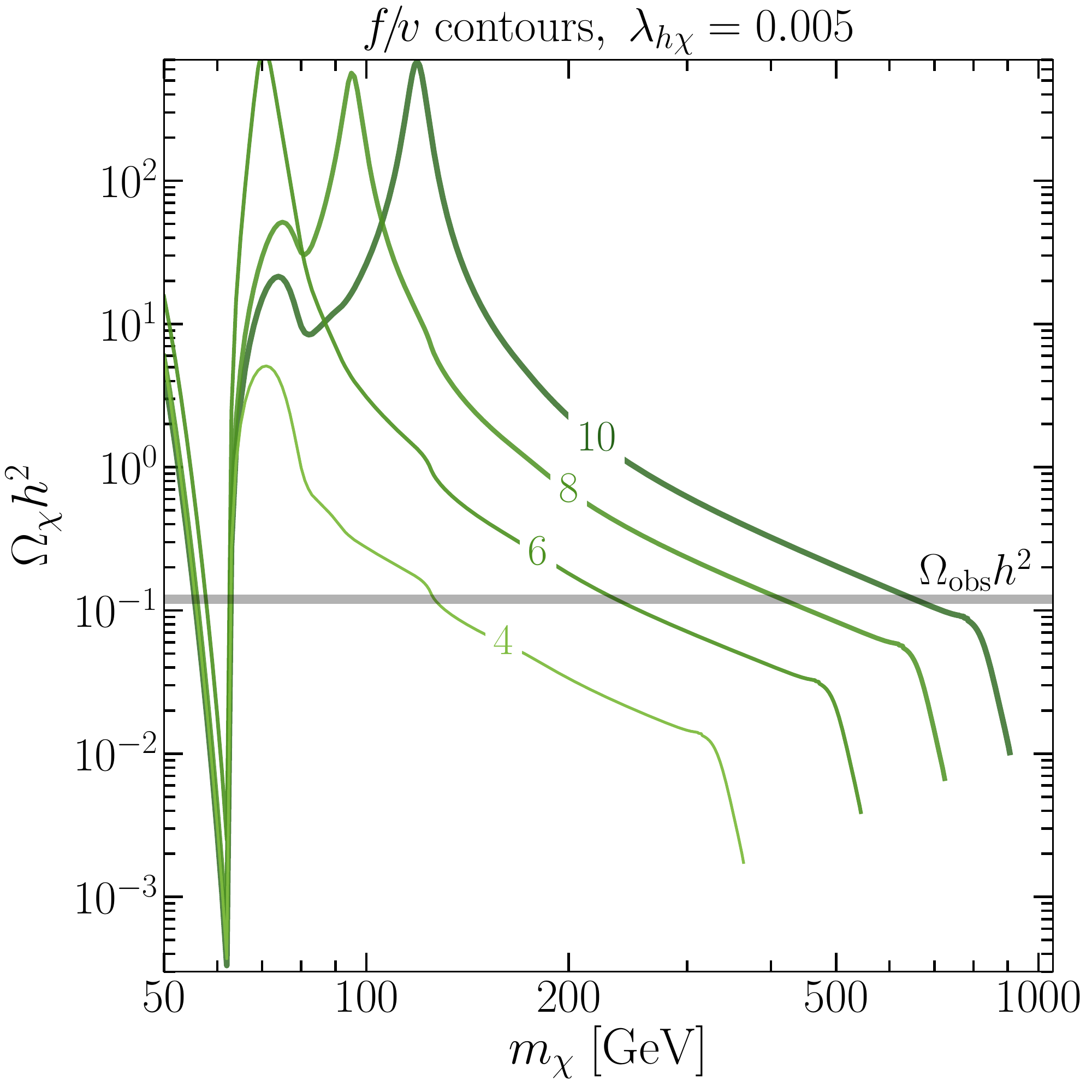} 
\hspace{-7pt}
\includegraphics[width=0.5\textwidth]{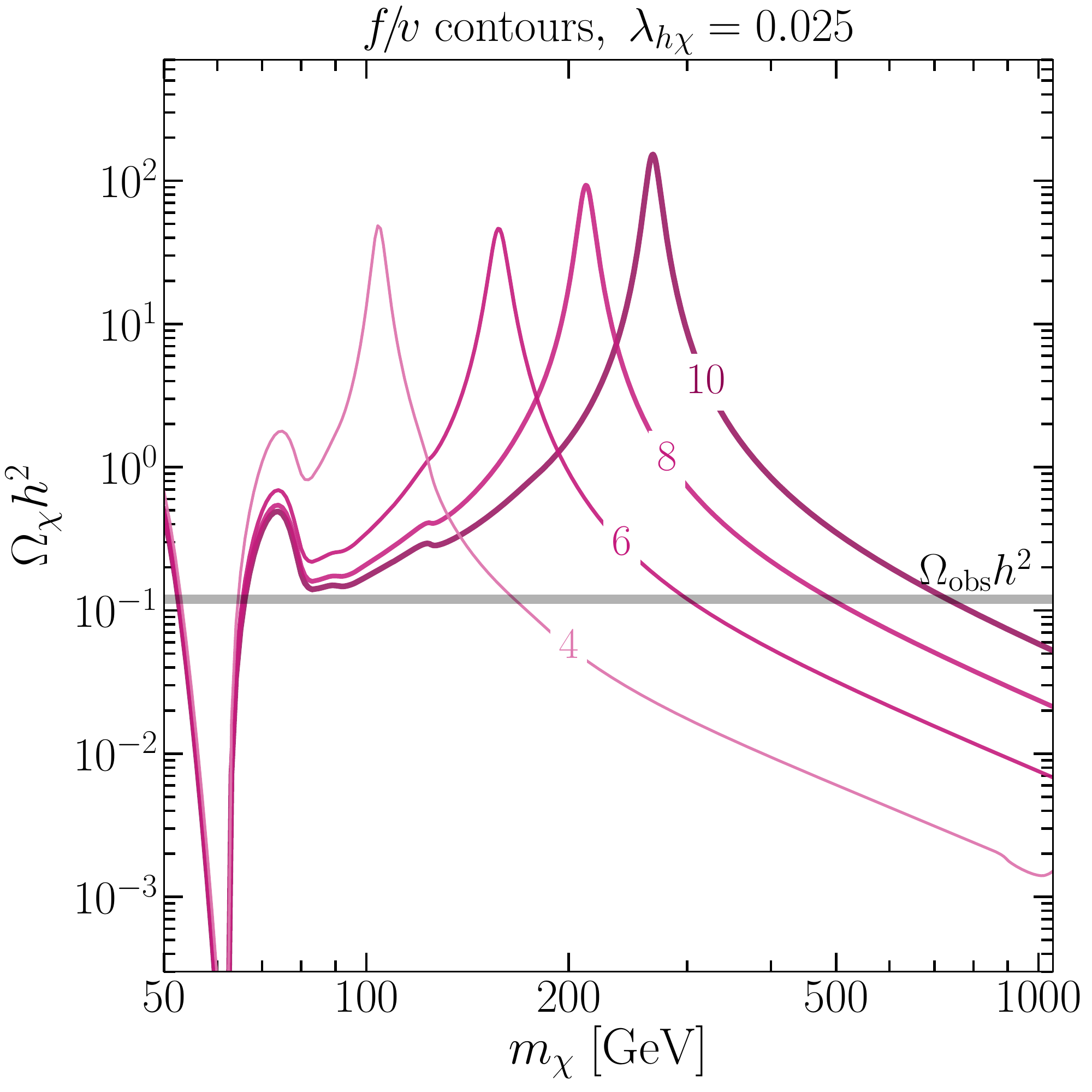} 
\caption{The left (right) plot shows the DM relic abundance $\Omega_{\chi} h^2$ as a function of DM mass $m_\chi$ for $\lambda_{h\chi}\!=\!0.005\, (0.025)$ with values of $f\!/\!v\!=\!4,6,8,10$. The gray line indicates observed relic abundance $\Omega_{\rm obs} h^2\!=\!0.12$.}
\label{fig:chi_relic}
\end{figure}

In \fig{fig:chi_relic}, we show the relic abundance $\Omega_\chi h^2$ for two benchmark values of $\lambda_{h\chi}\!=\!0.005$ and $0.025$ as a function of $m_\chi$ with fixed $f\!/\!v\!=\!4,6,8,10$. 
Notice that for masses below $50\gev$ the DM tends to be overproduced. This is because the thermal averaged cross-section in this region is directly proportional to the portal coupling $\lambda_{h\chi}$, which direct detection constrains to be relatively small (see below). 
On the other hand, for $m_\chi\! \sim\! m_h/2$ the relic abundance drops sharply due to the resonant enhancement of the Higgs portal cross-section. 
For DM masses $m_\chi^2\sim \lambda_{h\chi} f^2/2$ there is cancelation in the cross-section as a result the relic abundance enhances which produces the peaks in \fig{fig:chi_relic}.
For larger DM masses the cross section is proportional to $m_\chi^2/f^4$ and the relic density drops as DM mass increases. 

For the case $\lambda_{h\chi}\!=\!0.005$ (left-panel), the relic density curves terminate when the DM becomes heavier than the quirk states $\what X,\what Y$. These states are bound by the dark color force into quirky bound states, which then efficiently annihilate due to their electric charge, making them an unsuitable thermal DM candidate. 
There is also a sharp drop in the relic density at the end of each curve, which is due to an $s$-channel resonant enhancement of semi-annihilation processes, i.e. $\chi\what V\to \what T_\pm\to\what T_\pm {\rm SM}$, as shown in \fig{fig:chichi_semiann}. 
The semi-annihilation processes are only significant when $m_\chi\approx m_V\approx m_{\pm}/2$ and in most of the parameter space are inefficient as compared to the standard annihilation processes. 
Since the portal coupling $\lambda_{h\chi}$ is proportional to $r_Q=m_V^2/m_Q^2$ it can be reduced for relatively light vector-like quirks $\what V$. 
However, collider searches at the LEP and LHC put a lower bound these vector-like quirks, see \sec{s.QuirkySignals}. 

We see that the smallest mass that produces the correct DM thermal relic is near the Higgs resonance region, above $\sim\!50\gev$. This is fairly independent of $f\!/\!v$ and $\lambda_{h\chi}$. However, the largest DM masses which leads to correct relic abundance does depend on $f\!/\!v$ and $\lambda_{h\chi}$. Since naturalness prefers a smaller $f\!/\!v$ and $\lambda_{h\chi}$ is constrained by direct detection (see below), we find that restricting $f\!/\!v\!\leq\!10$ puts an upper bound of $m_\chi\lesssim1\tev$ for obtaining the correct relic.

%%%%%%%%%%%%%%%%%%%%
\subsection{Direct detection}
\label{s.direct_det}
%%%%%%%%%%%%%%%%%%%%
The WIMP DM scenario is being thoroughly tested by direct detection experiments. 
We here highlight the main features of our pNGB DM construction where direction detection null results are explained naturally. 

At tree-level the DM-nucleon interaction is only mediated by $t$-channel Higgs exchange. 
As discussed above, the DM-Higgs interaction has two sources: (i) the derivative coupling $\sim\! (\partial_\mu h)\partial_\mu(\chi^\ast\chi)/f^2$, and (ii) the portal coupling $\sim\! \lambda_{h\chi} h\chi^\ast\chi$. 
The strength of the derivative interaction in a $t$-channel process is suppressed by the DM momentum transferred, $t/f^2\!\sim\!(100\mev)^2/f^2\ll 1$. 
For all practical purposes we can neglect such interactions. 
Hence the only relevant interaction for direct detection is the portal coupling $\lambda_{h\chi}$.\footnote{There are 1-loop processes involving the quirk states and the electroweak  bosons which contribute to the DM-nucleon scattering. These processes are suppressed compared to tree-level, so we neglect them.}
In this case, the spin-independent DM-nucleon scattering cross-section $\sigma_{\chi N}^{\rm SI}$ can be approximated as (see e.g.~\cite{Frigerio:2012uc,Balkin:2017aep}), 
\beq
\sigma_{\chi N}^{\rm SI}\simeq \frac{f_N^2 m_N^4}{\pi m_h^4}\frac{\lambda_{h\chi}^2}{m_\chi^2}\approx 2.5\!\times\!10^{-46}\,{\rm cm}^2\left(\frac{\lambda_{h\chi}}{0.025}\right)^2\left(\frac{300\gev}{m_\chi}\right)^2,	\label{eq:sigmachin}
\eeq
where $m_N$ is the nucleon mass and $f_N\simeq 0.3$ encapsulates the Higgs-nucleon coupling. 
The current bound on the spin-independent DM-nucleon cross-section for mass range $\sim\![50,\,1000]$ GeV is by XENON1T with one ton-year of exposure time~\cite{Aprile:2018dbl}. For instance, the upper limit on the spin-independent DM-nucleon cross-section for DM mass $300\gev$ is $\sim\!2.5\!\times\!10^{-46}$ at $90\%$~C.L. 
From \eq{eq:sigmachin} it is clear that the $\sigma_{\chi N}^{\rm SI}$ is directly proportional to the square of the portal coupling $\lambda_{h\chi}^2$ and inversely proportional to the square of DM mass $m_\chi^2$. Hence to satisfy the direct detection constraints we either need to reduce the portal coupling $\lambda_{h\chi}$ or increase the DM mass. 

One feature of this minimal model is that $\lambda_{h\chi}$ is determined by a small number of low-energy parameters: the vector-like masses of the quirks, $m_V$ and $m_Q$. 
However, as noted above in \eq{eq:mQ}, the top partners quirk mass $m_Q\!=\!c_v \lambda_t f$ is fixed in terms of $f$ to obtain the correct Higgs mass. Hence, the free parameters are $m_\chi,f$, and $r_Q\!\equiv\!m_V^2/m_Q^2$. 
As discussed above, one can specify $f$ by requiring the correct DM relic abundance and $r_Q$ can be exchanged with $\lambda_{h\chi}$, which is constrained by direct detection.

\begin{figure}[t!]
\centering
\includegraphics[width=0.67\textwidth]{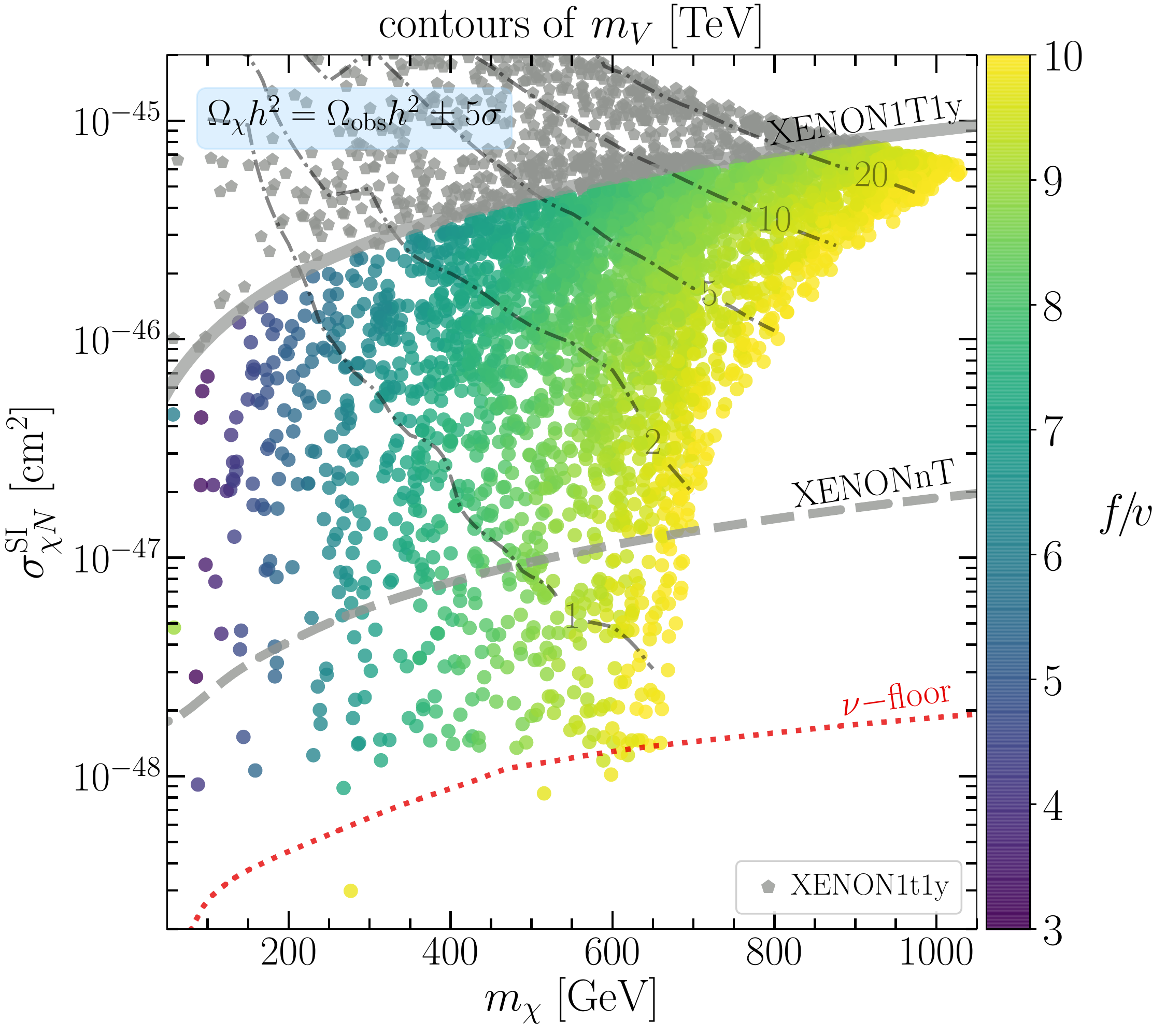} 
\caption{The DM-nucleon cross section as a function of DM mass $m_\chi$ with all points producing the observed relic abundance $\Omega_{\rm obs} h^2\!=\!0.12\pm0.0012$. The gray points above gray curve are excluded by current XENON1T bounds, whereas the colored points (corresponding to particular $f\!/\!v$ values) are allowed.}
\label{fig:chi_xsecchiN}
\end{figure}

In \fig{fig:chi_xsecchiN} we show the spin-independent DM-nucleon cross section $\sigma_{\chi N}^{\rm SI}$ as a function of DM mass $m_\chi$. 
We have performed a random scan of the parameter space for $f\!/\!v\!\in\![3,10]$ and $m_V\!\in\![m_\chi,4\pi f]$. 
The lower value of the $f\!/\!v\!=\!3$ choice is enforced by the SM Higgs coupling measurement and electroweak measurements data, while the upper value of $f\!/\!v\!=\!10$ limits the tuning to $\sim$1\%.
The lower value of $m_V$ makes sure that $\chi$ is the lightest state charged under $U(1)_D$. 
All the points shown in the plot correspond to the correct relic abundance $\Omega_\chi h^2\!=\!\Omega_{\rm obs} h^2\!\pm 5\sigma$, where $\Omega_{\rm obs} h^2\!=\!0.12\pm0.0012$ is the observed DM relic density as measured by the Planck satellite~\cite{Aghanim:2018eyx}. 
The gray (pentagon) points above the gray line are excluded by the XENON1T~\cite{Aprile:2018dbl}. 
All the colored points (color barcoded with $f\!/\!v$) are allowed by the current XENON1T constraint.
The dashed gray line indicates the expected XENONnT bound~\cite{Aprile:2018dbl} which covers much of the more natural parameter space. However, there are points allowed below this bound above the so-called neutrino-floor (red dotted), which could be discovered by next generation detectors, e.g. LZ~\cite{Akerib:2018dfk} and DARWIN~\cite{Aalbers:2016jon}. 
%

%%%%%%%%%%%%%%%%%%%%
\subsection{Indirect detection\label{s.indirect_det}}
%%%%%%%%%%%%%%%%%%%%
We now turn to indirect detection. There are a variety of experiments searching for DM annihilations in the Milky Way galaxy and nearby dwarf galaxies, which are assumed to be dominated by DM. 
The typical signals of DM annihilation to the SM particles leads to gamma-rays, gamma-lines, and an excess of secondary products like antipositrons and antiprotons in cosmic-rays (CR). In particular, the experimental data can be used to put upper bounds on the various annihilation channels, including $WW,ZZ,hh,t\bar t, b\bar b, \tau^+\tau^-,\cdots$.  
In our model the DM dominantly annihilates into $WW,hh,ZZ,t\bar t$ final states. 
We calculate the present day DM thermal averaged annihilation to the SM particles by using \texttt{micrOMEGAs}~\cite{Belanger:2018mqt}. 
We find that the DM thermal annihilation cross-section is $\langle\sigma v\rangle \approx 2.2\times 10^{-26}\,\mathrm{cm^3/s}$ for parameter values that produce the correct relic abundance.
The fraction of annihilation cross-section to $W^+W^-$ is $\sim\!45\%$ and $hh/ZZ\sim\!25\%$ for $m_\chi\!\gtrsim \!m_h$. Whereas the branching fraction is dominantly into $W^+W^-$ for $m_\chi\in[m_W,m_h]$. 

\begin{figure}[t!]
\centering
\includegraphics[width=0.67\textwidth]{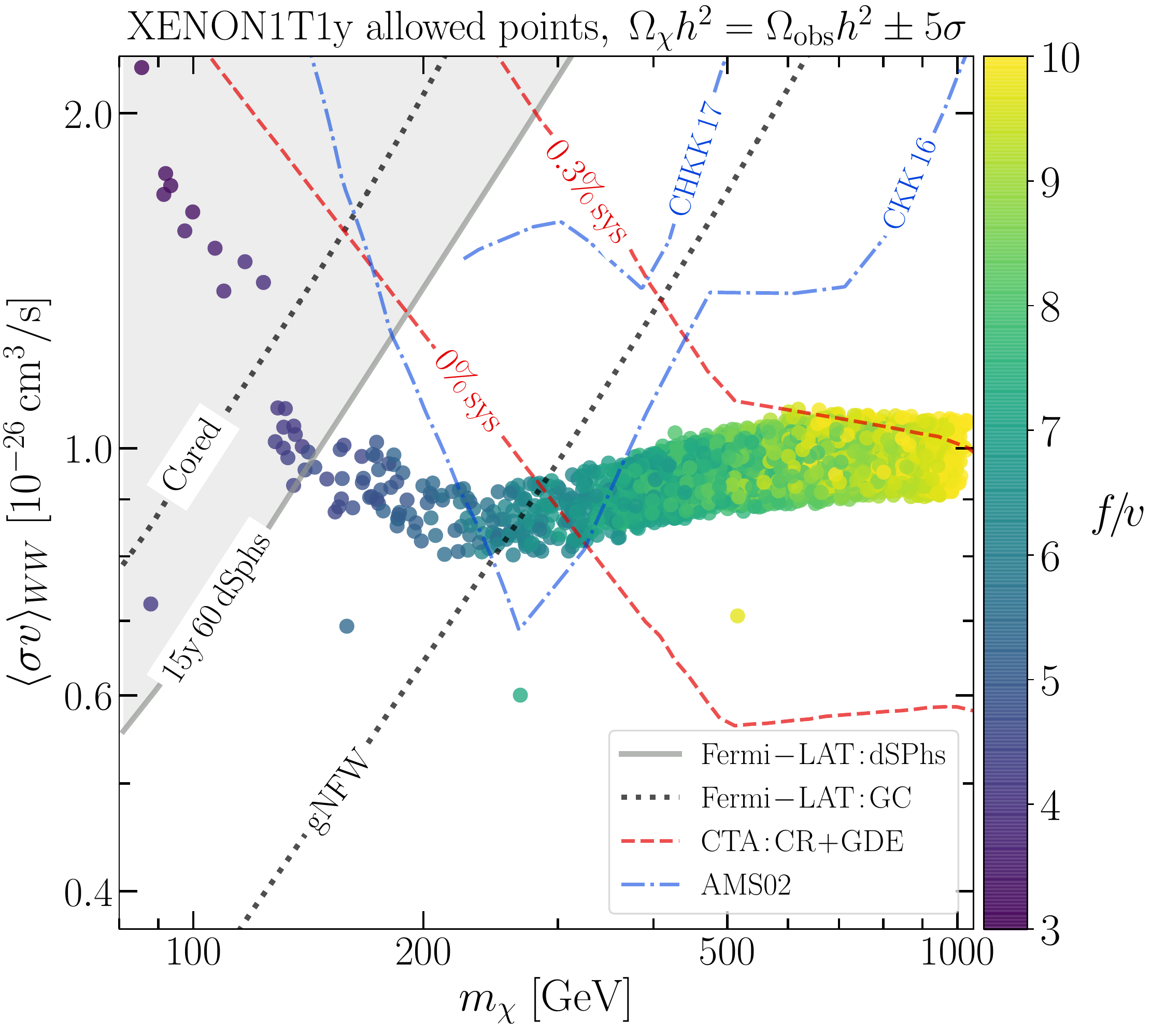} 
\caption{A parameter scan with all the points producing correct relic abundance in the $m_\chi$ vs $\langle\sigma v\rangle_{WW}$. Each point (color coded for different $f\!/\!v$ values) is allowed by the XENON1T1y direct detection experiment. The gray shaded region is the robust projected 95\% C.L. sensitivity of \texttt{Fermi-LAT} with 15 years data and 60 dSphs. The dashed curves represents the \texttt{CTA} projected 95\% C.L. sensitivity with background only (0\% systematics) and 0.3\% systematics, whereas the dotted and dash-dotted curves are relatively less robust 95\% C.L. upper bounds due to \texttt{Fermi-LAT} and \texttt{AMS-02}, respectively.}
\label{fig:chi_sigmav}
\vspace{-10pt}
\end{figure}

In \fig{fig:chi_sigmav} we show the DM annihilation cross section to $W^+W^-$, $\langle\sigma v\rangle_{WW}$, in units of $[10^{-26}~\mathrm{cm^3/s}]$ as function of $m_\chi$. 
All the data points in this figure produce correct DM relic abundance and satisfy the XENON1T direct detection constraint. 
Because these points have $m_\chi\!>\!m_W$, the most dominant annihilation channels are the $WW,ZZ,hh$. In the following we summarize the most sensitive indirect detection probes in the mass range of interest. 

\paragraph{Gamma-rays:} 
The most robust indirect detection bounds are due to \texttt{Fermi-LAT}~\cite{Ackermann:2015zua} and \texttt{Fermi-LAT+DES}~\cite{Fermi-LAT:2016uux} with six years of data from 15 and 45 DM dominated dwarf spheroidal galaxies (dSphs), respectively. 
Theses constraints are considered robust because the uncertainties associated with propagation of gamma rays are relatively small. 
The \texttt{Fermi-LAT} results~\cite{Ackermann:2015zua} provide upper-limits on the DM thermal annihilation cross section into several SM final states including $WW, b\bar b, \tau^+\tau^-$, whereas, the updated analysis \texttt{Fermi-LAT+DES}~\cite{Fermi-LAT:2016uux} only includes the $b\bar b$ and $ \tau^+\tau^-$ channels. 
These bounds do not constraint any of the parameter space allowed by the direct detection. 
However, \texttt{Fermi-LAT} has provided expected 95\% C.L. upper-limits for the DM thermal annihilation into $b\bar b$ and $\tau^+\tau^-$ channels with 15 years of data and 60 dSPhs~\cite{Charles:2016pgz}. 
One can interpolate the projected upper-limit from the $\langle \sigma v\rangle_{b\bar b}$ to $\langle \sigma v\rangle_{WW}$ by a simple rescaling $\langle \sigma v\rangle_{WW}\simeq1.33\langle \sigma v\rangle_{b\bar b}$ in the DM mass range of our interest. 
In \fig{fig:chi_sigmav} we show the projected 95\% C.L. sensitivity on $\langle \sigma v\rangle_{WW}$ by \texttt{Fermi-LAT} with 15 years and 60 dSPhs by the solid (gray) curve. 
This sensitivity sets a lower-limit on the  DM mass $m_\chi\gtrsim 150\gev$.

A very recent analysis~\cite{Abazajian:2020tww} of \texttt{Fermi-LAT} observations of the Galactic Center (GC) has led to stringent constraints on WIMP DM mass up to $\sim\!300\gev$. 
In \fig{fig:chi_sigmav} we show 95\% C.L. upper bound on $\langle \sigma v\rangle_{WW}$ as dotted (black) curves due to two DM profiles: a generalized NFW (gNFW) profile and a cored profile that smoothly matches on to a NFW profile while conserving mass. 
The upper limit on the thermal annihilation for each DM profile in \fig{fig:chi_sigmav} is least constraining when variations of the DM profiles and the systematic uncertainties associated with different Galactic Diffuse Emission (GDE) templets are taken into account, see~\cite{Abazajian:2020tww} for further technical details. 
The \texttt{Fermi-LAT} GC 95\% C.L. upper limit with a gNFW profile excludes our DM up to masses $\sim\!300\gev$. 
Assuming a cored profile, however, weakens the bound to $m_\chi\gtrsim100\gev$. While the GC constraints are highly sensitive to the DM profiles and the GDE templets, they still provide an important complimentary DM probe, and with more data these uncertainties will be reduced. 

\paragraph{Cosmic-rays:}
The flux of antipositrons and antiprotons in the cosmic-rays (CR) provides another indirect probe of DM annihilation in the Galaxy. 
In particular recent precise~\texttt{AMS-02} CR antiproton flux data~\cite{Aguilar:2016kjl} has led to strong constraints on the DM thermal annihilation. 
In Refs.~\cite{Cuoco:2016eej,Cui:2016ppb} the \texttt{AMS-02} antiproton flux data was used to put stringent constraints on DM with masses in range $[150,1000]\gev$, see also~\cite{Arina:2019tib} for a recent global fit analysis of pNGB DM. 
The \texttt{AMS-02} 95\% C.L. exclusion constraint on $\langle \sigma v\rangle_{WW}$ as obtained by CKK~\cite{Cuoco:2016eej} is shown in \fig{fig:chi_sigmav} as dash-dotted (blue) curve. This constraint excludes most of the data points between DM masses $m_\chi\in[225,375]\gev$. However, these constraint has large systematic uncertainties, mainly due to CR propagation and diffusion parameters~\cite{Cuoco:2016eej}. The updated analysis by (CHKK)~\cite{Cuoco:2017iax} reveals a weaker constraint in the $W^+W^-$ channel, which is also given by a dash-dotted (blue) curve. Even though the updated \texttt{AMS-02} analysis does not constrain our model, future \texttt{AMS} CR antiproton data are likely to. Another future CR experiment is the Cherenkov Telescope Array (CTA) which is expected to be sensitive to large DM masses~\cite{CTA:2018}. In \fig{fig:chi_sigmav} we show the projected sensitivity of CTA for DM annihilation to $W^+W^-$ with GDE Gamma model of Ref.~\cite{Gaggero:2017jts}, as a dashed (red) curve, for two assumptions of systematic error. The most optimistic implies that CTA will probe DM masses above $\sim\!300\gev$, though this is quickly weakened when systematic errors are included.

%%%%%%%%%%%%%%%%%%%%%%%%%%%%%%%%%%%
\section{Conclusion\label{s.conclusion}}
\begin{figure}[t!]
\centering
\includegraphics[width=0.67\textwidth]{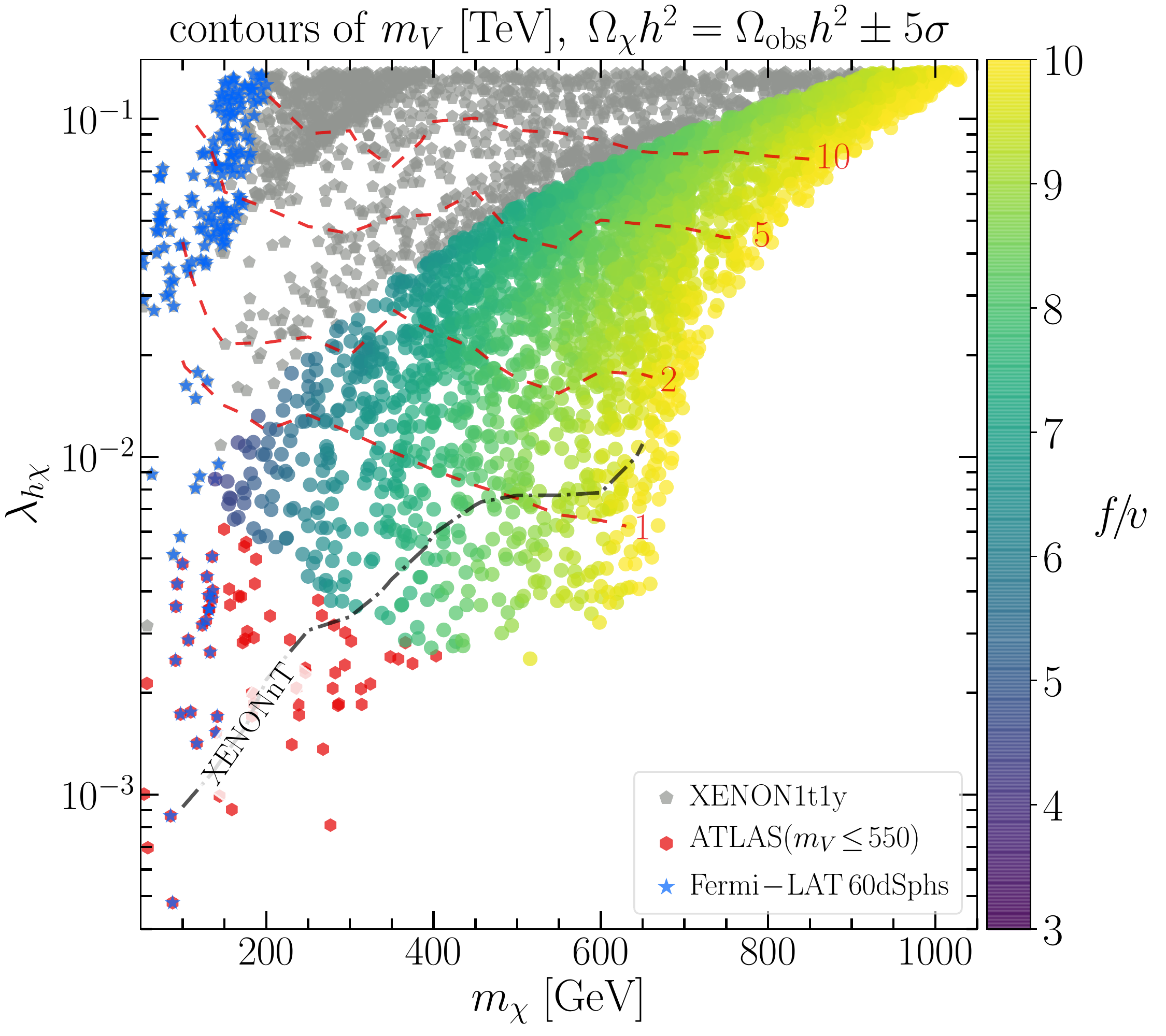}
\caption{A parameter scan with all points producing the correct relic abundance in the $m_\chi$ vs $\lambda_{h\chi}$. The colored (gray hexagon) points allowed (excluded) by the XENON1T1y direct detection experiment. The allowed points are color coded for different $f\!/\!v$ values. The red (pentagon) points are excluded by the ATLAS search of quirky states with $m_V\lesssim 550\gev$, while the indirect detection projected 95\% C.L. sensitive points due to \texttt{Fermi-LAT} with 15 years and 60 dSPhs are marked as blue (stars). The dashed (red) contours represent different values of $m_V$ [TeV], whereas the dash-dotted (black) curve gives the projected sensitivity of XENONnT.}
\label{fig:chi_lamhchi}
\vspace{-10pt}
\end{figure}

We have outlined a framework in which the Higgs and a scalar DM candidate arise pNGBs of a broken global symmetry. Because the symmetry partners of the top quark do not carry SM color, the induced scalar potential between the Higgs and the DM, which is UV insensitive, allows for improved fine-tuning and simultaneously explains null results for WIMP DM searches. The quantitative success of this framework is summarized by Fig.~\ref{fig:chi_lamhchi} in the $m_\chi$ vs $\lambda_{h\chi}$ plane with the color of scanned points corresponding to values of $f\!/\!v\in[3,10]$. This corresponds to fine-tuning in the model of about 10\% to 1\%, respectively. 

The phenomenology can be specified by the DM mass $m_\chi$, the global symmetry breaking scale $f$, and the vector-like mass $m_V$ of the quirky fermions, which is the source of breaking the $\chi$ shift symmetry. As shown in Sec.~\ref{sec:ScalarPotential} we can trade $m_V$ for $\lambda_{h\chi}$. Hence, the three free parameters of the model are $m_\chi,f\!/\!v$, and $\lambda_{h\chi}$. 

The points in \fig{fig:chi_lamhchi} scan in $m_\chi\!\sim\![50,1000]\gev$ and $\lambda_{h\chi}\!\sim\![0.2,0.0005]$ while required to produce the correct relic abundance $\Omega_\chi h^2=0.12\pm5(0.012)$. The gray (pentagon) points are excluded at 90\% C.L. by the direct detection experiment XENON1T with one year exposure time~\cite{Aprile:2018dbl}. 
Future direct detection XENONnT 90\% C.L. reach is overlaid as the dash-dotted (black) curve, which would cover much of the allowed parameter space. Next generation experiments that will descend toward the neutrino floor will fully explore this framework.

The next most stringent constraint is due to the LHC bound on the vector-like mass $m_V\gtrsim550\gev$ of the quirky fermions $\what X,\what Y$ as shown in \fig{f.WWsearch}. This limit from the ATLAS collaboration search for dilepton resonances with $\mathrm{139~fb^{-1}}$ data is due to the annihilation of quirks $\what V\what V$ to $\ell^+\ell^-$. We show the bound in \fig{fig:chi_lamhchi} as red (hexagon) points. Since the portal coupling $\lambda_{h\chi}$ is proportional to $m_V^2$, the lower-bound on $m_V$ translates to a DM mass and $f\!/\!v$ dependent lower-bound on $\lambda_{h\chi}$.  
We have also shown dashed (red) contours of $m_V=1\tev$ to $10\tev$ which shows how future LHC runs may be able to discover quirks in much of the natural parameter space. The complementarity between collider and direct detection could lead to both discovery and confirmation of this construction in the coming years, or its exclusion. 

In \fig{fig:chi_lamhchi} we also show how indirect detection gamma-rays 95\% C.L. constraints from the \texttt{Fermi-LAT} 15 years with 60 dSphs as blue (star) points. 
This puts a lower limit on the DM mass $m_\chi\gtrsim150\gev$. 
We have not shown in this plot the indirect detection constraints from the cosmic-rays experiments \texttt{AMS-02} because of their large systematic uncertainties. 
However, in the future such uncertainties may be reduced, allowing experiments like \texttt{AMS} and \texttt{CTA}) to provide another complementary probe, and hopefully discovery, of this model. 
%%%%%%%%%%%%%%%%%%%%%%%%%%%%%%%%%%%

In summary, this framework of WIMP dark addresses the hierarchy problem without colored symmetry partners, and consequently is only tuned at the 10\% level while agreeing with all experimental bounds. However, existing experiments will soon be able to discover or exclude these more natural realizations of the model. After the searches of the HL-LHC run and next generation direct detection experiments models with fine tuning at or better than 1\% may be thoroughly probed.

%%%%%%%%%%%%%%%%%%%
\section*{Acknowledgements}
We thank Zackaria Chacko for encouraging this study. We also thank Matthew Low and Roni Harnik for enlightening discussions along with Lingfeng Li and Ennio Salvioni for assistance with quirk dynamics. 
A.A. and S.N. are supported by FWO under the EOS-be.h project no. 30820817 and Vrije Universiteit Brussel through the Strategic Research Program ``High Energy Physics''. 
C.B.V is supported in part by NSF Grant No. PHY-1915005 and in part by Simons Investigator Award  \#376204.

%%%%%%%%%%%%%%%%%%%%%%%%%%%%%%%%%%%
\appendix
%%%%%%%%%%%%%%%%%%%%%%%%%%%%%%%%%%%
%%%%%%%%%%%%%%%%%%%%%%%%%%%%%%%%%%%
\section{$SO(7)$ Generators}
\label{app:gens}
%%%%%%%%%%%%%%%%%%%%%%%%%%%%%%%%%%%
In this appendix we collect all the relevant details. The $SO(7)$ generators in the fundamental representation can be written as, 
\begin{align}
T_{i j}^{a_{L, R}} &=-\frac{i}{2}\left[\frac{1}{2} \epsilon^{a b c}\left(\delta_{i}^{b} \delta_{j}^{c}-\delta_{j}^{b} \delta_{i}^{c}\right) \pm\left(\delta_{i}^{a} \delta_{j}^{4}-\delta_{j}^{a} \delta_{i}^{4}\right)\right], 	&a_{L,R}&=1,2,3,	\\ 
T_{ij}^{a b}&=-\frac{i}{\sqrt{2}}\left(\delta_{i}^{a} \delta_{j}^{b}-\delta_{j}^{a} \delta_{i}^{b}\right), \quad &b=5,6 ; ~a&=1, \ldots, b-1,	\\
T_{i j}^{\hat{a}} &=-\frac{i}{\sqrt{2}}\left(\delta_{i}^{\hat{a}} \delta_{j}^{7}-\delta_{j}^{\hat{a}} \delta_{i}^{7}\right), 	&\hat a&=1,\ldots,6,
\end{align}
where $i,j\!=\! 1, \ldots,7$. We have chosen the normalization $\text{Tr}\left[T^aT^b \right]=\delta^{ab}$. The unbroken generators $T_{ij}^{a_{L,R}},T_{ij}^{a b}$ correspond to the $SO(6)$, whereas the broken generators $T_{i j}^{\hat{a}}$ correspond to the $SO(7)/SO(6)$ coset. Note that $T_{ij}^{a_{L,R}}$ correspond to the custodial $SO(4)_C \cong S U(2)_{L} \times S U(2)_{R}$ subgroup of $SO(6)$.

%%%%%%%%%%%%%%%%%%
\section{Feynman rules and Quirk Processes\label{app:FeyQuirk}}

\begin{table}[h!]
\centering
\includegraphics[width=\textwidth]{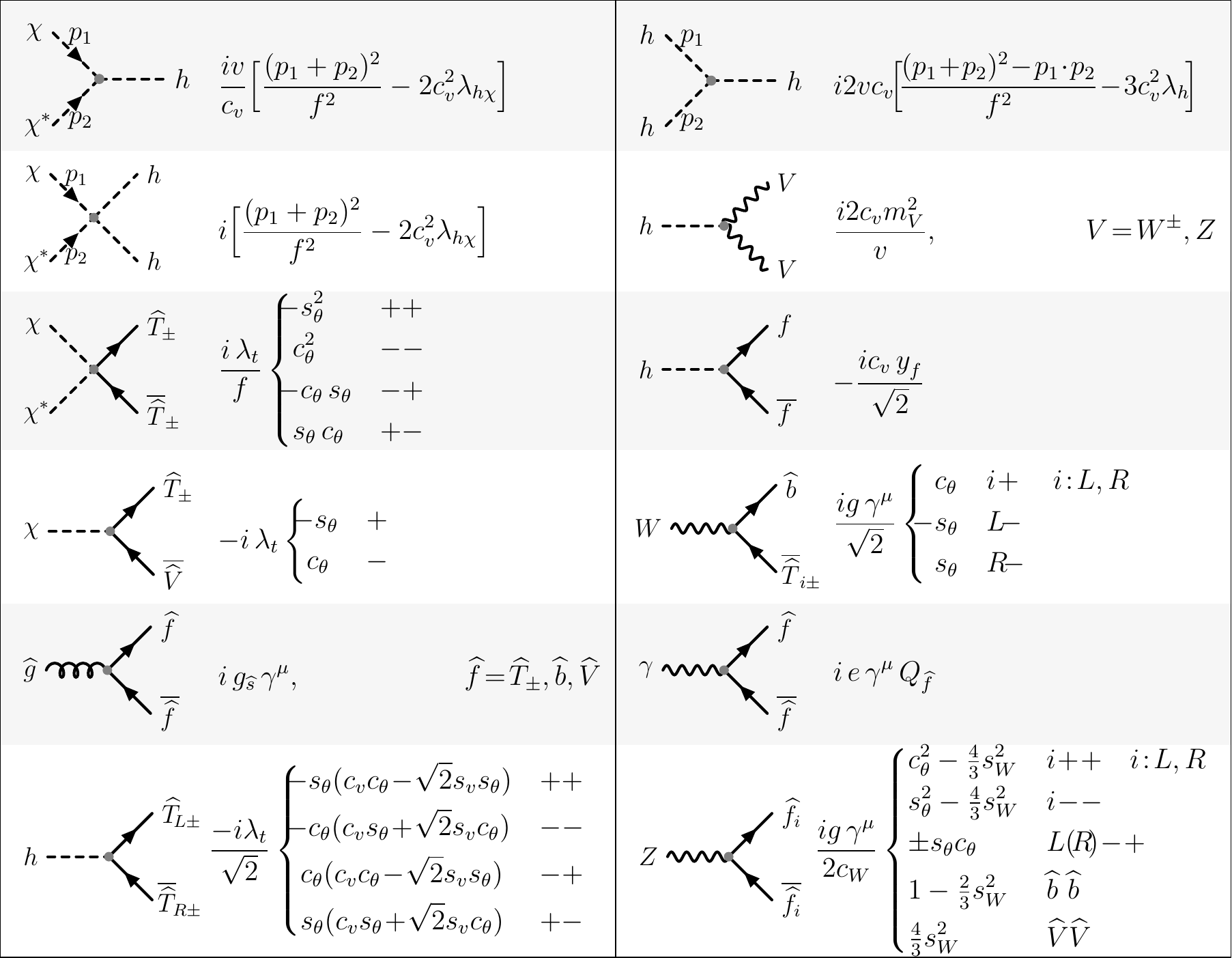} 
\caption{Some of the most relevant Feynman rules of our model are listed in this table, see the text for the corresponding notation.}
\label{tab:feynrules}
\end{table}
In this appendix we record formulae for quirk production and decay widths. The relevant Feynman rules are given in Table~\ref{tab:feynrules}. The decays are typically similar to the results \cite{Barger:1987xg,Fok:2011yc}, using the methods outlined in~\cite{Kuhn:1979bb,Guberina:1980dc}. The couplings of the $Z$ to fermions are taken to be
\begin{equation}
\frac{g}{2 c_W}\gamma^\mu(v_i- a_i \gamma_5),\label{e.ZfermionCoupling}
\end{equation}
where $c_W\equiv \cos\theta_W$. For convenience we define the following
\begin{equation}
R_i=\frac{m_i^2}{M^2}, \ \ \ \  \beta_{ij}=\sqrt{1-2(R_i+R_j)+(R_i-R_j)^2},
\end{equation}
where $M$ is the mass of the relevant bound state. The number of colors in the quirk confining group is $N_{\what{c}}$. 

We calculate the cross section $pp\to Z,\gamma\to \overline{f}f$ from the quark $q$ initiated partonic cross section $\tilde{\sigma}$ into a quirk $Q$ pair by
\beq
\sigma(pp\to QQ)(s)=\sum_q\int_\frac{4m_Q^2}{s}d\tau L_{\overline{q}q}\tilde{\sigma}(\overline{q}q\to \overline{Q}Q)(\tilde{s}=\tau s),
\eeq
where 
\beq
L_{\overline{q}q}(\tau)=\int^1_\tau\frac{dx}{x}\left[f_q(x)f_{\overline{q}}\left( \frac{\tau}{x}\right) +f_q\left(\frac{\tau}{x}\right)f_{\overline{q}}(x)\right],
\eeq
is defined in terms of the MSTW2008 PDFs~\cite{Martin:2009iq} $f_q(x)$, we take the factorization scale to be $\sqrt{\tilde{s}}/2$.

Because the quirk states decays from all $\ell>0$ states are strongly suppressed~\cite{Kang:2008ea} we only consider decays of the singlet ${}^1\! S_0$ and triplet ${}^3\! S_1$ states. Each of these decay widths depends on the radial wavefunction $R(0)$ of the quirk bound state. This factor is nonperturbative and not exactly known, so we simply give each decay width in units of the unknown factor $\left|R(0) \right|^2$. 

The neutral states are composed of fermionic quirks $Q$ with mass $m_Q$. In this case the $Z$ couplings are labeled $v_Q$ and $a_Q$, and the electric charge is denoted $Q_Q$.\footnote{This introduces a relative factor of two compared to the $Z$ couplings used by \cite{Barger:1987xg,Fok:2011yc}.} The mass is denoted $m_Q$ and we take the meson mass to be $M$, which for heavy constituents is approximately $2m_Q$.

We begin with decays to fermion pairs. These fermions have $Z$ couplings $v_f$ and $a_f$ as well as electric charge $Q_f$. They also come in $N_c$ colors. The decays to $\overline{f} f$ are, 
\begin{align}
\Gamma({}^1\! S_0\to \overline{f}f)=&\frac{2N_{\what{c}}N_c\alpha_W^2a_f^2a_Q^2m_f^2m_Q^2}{c_W^4m_Z^4M^2}\beta_{ff},\\
\Gamma({}^3\! S_1\to \overline{f}f)=&\frac{N_{\what{c}}N_c\alpha_W^2}{12M^2}\beta_{ff}\left[(1+2R_f)\left(4s_W^2Q_QQ_f+\frac{v_Qv_f}{c_W^2(1-R_Z)} \right)^2 +\frac{v_Q^2a_f^2\beta_{ff}^2}{c_W^4(1-R_Z)^2}\right],
\end{align}
where  $\alpha_W\equiv g^2/(4\pi)$. Next, we turn to decays into $Z\gamma$, 
\begin{align}
\Gamma({}^1\! S_0\to Z\gamma)=&\frac{8N_{\what{c}}\alpha_W\alpha Q_Q^2 v_Q^2}{c_W^2M^2}(1-R_Z),\\
\Gamma({}^3\! S_1\to Z\gamma)=&\frac{8N_{\what{c}}\alpha_W\alpha Q_Q^2a_Q^2m_Q^2}{3c_W^2m^2_ZM^2}(1-R_Z^2).
\end{align}
The decays to $ZZ$~\footnote{Note the erratum of \cite{Barger:1987xg} in reference to $\Gamma({}^1\! S_0\to ZZ)$ and $\Gamma({}^3\! S_1\to Zh)$. In addition, the $\Gamma({}^1\! S_0\to f\bar{f})$ depends on the axial coupling of the ${}^1\! S_0$ constituents to the $Z$, as clarified in \cite{Fok:2011yc}.},
\begin{align}
\Gamma({}^1\! S_0\to ZZ)=&\frac{N_{\what{c}}\alpha_W^2(v_Q^2+a_Q^2)^2}{4M^2c_W^4(1-2R_Z)^2}\beta_{ZZ}^3,\\
\Gamma({}^3\! S_1\to ZZ)=&\frac{N_{\what{c}}\alpha_W^2v_Q^2a_Q^2}{3c_W^2M^2R_Z(1-2R_Z)^2}\beta_{ZZ}^5.
\end{align}
Next, to $Zh$,
\begin{align}
\Gamma({}^1\! S_0\to Zh)=&\frac{N_{\what{c}}\alpha_W^2a_Q^2M^2}{16m_Z^4c_W^4}\beta_{Zh}^3,\\
\Gamma({}^3\! S_1\to Zh)=&\frac{N_{\what{c}}\alpha_W^2v_Q^2\beta_{Zh}}{12c_W^2M^2m_W^2}\left\{\left[2+\frac{1}{4R_Z}\left(1+R_Z-R_h \right)^2 \right]\left[\frac{2m_QR_Z}{1-R_Z}-\frac{v\lambda_Q(1+R_Z-R_W^2)}{1-R_Z-R_h} \right]^2\right.\nonumber\\
&\left.+\frac{v\lambda_Q\beta_{Zh}^2(1-R_h+R_Z)}{2R_Z(1-R_Z-R_h)}\left[\frac{2m_QR_Z}{1-R_Z}-\frac{v\lambda_Q(1+R_Z-R_W^2)}{1-R_Z-R_h} \right]+\frac{\beta_{Zh}^4v^2\lambda^2_Q}{4R_Z(1-R_Z-R_h)^2} \right\},
\end{align}
where $\lambda_Q$ is the Yukawa coupling of the quirks to the Higgs. Finally, to $h\gamma $,
\begin{align}
\Gamma({}^1\! S_0\to h \gamma )=&0,\\
\Gamma({}^3\! S_1\to h\gamma )=&\frac{N_{\what{c}}\alpha Q_Q^2\lambda^2_Q(1-R_h)}{3\pi M^2}.
\end{align}
One might expect decays to scalar pairs like $hh$ and, in the case of the $\widehat{X}$ and $\widehat{Y}$ quirks, $\chi\chi^\ast$. However, $CP$ and angular momentum conservation forbid such decays from the $s$-wave states, though higher angular momentum states do allow these decays. 

We now turn to decays into $W^+W^-$. We label the $SU(2)_L$ partner of $Q$ by $q$, with  mass $m_q$ etc. The $W$ couplings $v_W$ and $a_W$ are defined by the interaction
\beq
\frac{g}{2\sqrt{2}}\gamma^\mu(v_W-a_W\gamma^5).
\eeq
We note that this decay depends upon the electric charge of particle that makes up the bound state in a nontrivial way. This is due to the diagrams related to the $t$- or $u$-channel exchange of the $SU(2)_L$ partner of the particle making up the bound state. Mesons made by a quirk with positive charge involve a different diagram than those with negative charge. None of these subtleties affect the singlet case, but we do distinguish the triplet cases as ${}^3\! S_1^{(+,-)}$, where the superscript denotes whether the quirk has positive or negative electric charge. The decays to $W^+W^-$ are
\begin{align}
\Gamma({}^1\! S_0\to W^+W^-)&=\frac{N_{\what{c}}\alpha_W^2(v_W^2+a_W^2)^2}{8M^2(1+4R_q-4R_W)^2}\beta_{WW}^3,\\
\Gamma({}^3\! S_1^{(\pm)}\to W^+W^-)&=\frac{N_{\what{c}}\alpha_W^2\beta_{WW}^3}{192R_W^2M^2}\left\{\frac{16v_W^2a_W^2\beta_{WW}^2}{(1+4R_q-4R_W)^2}-6\frac{R_W}{m_Q^2}\left[\frac{v_W^2(m_Q-m_q)+a_W^2(m_Q+m_q)}{1+4R_q-4R_W} \right]^2\right.\nonumber\\
&+(1+10R_W)\left[4Q_Qs_W^2+\frac{2v_Q}{1-R_Z}\mp\frac{m_q}{m_Q}\frac{v_W^2-a_W^2}{1+4R_q-4R_W} \right]^2\nonumber\\
&\left. +2R_W(5+6R_W)\left[4Q_Qs_W^2+\frac{2v_Q}{1-R_Z}\mp\frac{v_W^2+a_W^2}{1+4R_q-4R_W} \right]^2 \right\}.
\end{align}

We also record the decays involving hidden gluons. These are taken from \cite{Cheung:2008ke}.
\begin{align}
\Gamma({}^1\! S_0\to \widehat{g}\,\widehat{g})=&\frac{N_{\what{c}}^2-1}{N_{\what{c}} M^2}\widehat{\alpha}_s^2,\\
\Gamma({}^3\! S_1\to \widehat{g}\,\widehat{g}\,\widehat{g})=&\frac{(N_{\what{c}}^2-1)(N_{\what{c}}^2-4)(\pi^2-9)}{9\pi N_{\what{c}}^2M^2}\widehat{\alpha}_s^3,\\
\Gamma({}^3\! S_1\to \gamma \widehat{g}\,\widehat{g})=&\frac{4Q_f^2(N_{\what{c}}^2-1)(\pi^2-9)}{3\pi M^2N_{\what{c}}}\alpha \widehat{\alpha}_s^2,
\end{align}
where we have denoted the hidden sector strong coupling by $\widehat{\alpha}_s$. Finally, the singlet state can also decay to photons
\beq
\Gamma({}^1\! S_0\to \gamma\gamma)=\frac{4N_{\what{c}}Q_Q^4}{M^2}\alpha^2.
\eeq

%%%%%%%%%%%%%%%%%%%

\bibliography{bib_so7nn}{}
\bibliographystyle{aabib}

%%%%%%%%%%%%%%%%%%
\end{document}